\documentclass[journal]{IEEEtran}
\ifCLASSINFOpdf
\else
\fi
\usepackage{amsmath}
\usepackage{amssymb}
\usepackage{graphicx}
\usepackage{multirow}
\usepackage{xcolor}

\newcommand{\prop}{PANDA}
\usepackage{hhline}
\usepackage[normalem]{ulem}
\useunder{\uline}{\ul}{}
\usepackage{tabularx}
\begin{document}
%
% paper title
% Titles are generally capitalized except for words such as a, an, and, as,
% at, but, by, for, in, nor, of, on, or, the, to and up, which are usually
% not capitalized unless they are the first or last word of the title.
% Linebreaks \\ can be used within to get better formatting as desired.
% Do not put math or special symbols in the title.
\title{Polarization Aware Deep Learning Framework for SAR Despeckling}
%
%
% author names and IEEE memberships
% note positions of commas and nonbreaking spaces ( ~ ) LaTeX will not break
% a structure at a ~ so this keeps an author's name from being broken across
% two lines.
% use \thanks{} to gain access to the first footnote area
% a separate \thanks must be used for each paragraph as LaTeX2e's \thanks
% was not built to handle multiple paragraphs
%

\author{Gianpaolo Passarello,~\IEEEmembership{Student Member,~IEEE,}
        Sergio Vitale,~\IEEEmembership{Member, IEEE,} Hossein Aghababaei,~\IEEEmembership{Senior Member, IEEE,} Giampaolo Ferraioli,~\IEEEmembership{Senior Member, IEEE,}
        Vito Pascazio,~\IEEEmembership{Life~Fellow,~IEEE}
    \thanks{Gianpaolo Passarello is with the Dipartimento di Ingegneria, Università degli Studi di Napoli “Parthenope,” 80143 Naples, Italy (e-mail: gianpaolo.passarello@phds.uniparthenope.it).}% <-this % stops a space
    \thanks{Sergio Vitale and Vito Pascazio are with the Dipartimento di Ingegneria, Università degli Studi di Napoli “Parthenope,” 80143 Naples, Italy, and also with the National Inter-University Consortium for Telecommunications (CNIT), 80126 Naples, Italy (e-mail: sergio.vitale@uniparthenope.it; vito.pascazio@uniparthenope.it).}%
    \thanks{Hossein Aghababaei is with the Department of Earth Observation Science, Faculty of Geo-Information Science and Earth Observation,University of Twente, Enschede 7514AE, The Netherlands (e-mail:h.aghababaei@utwente.nl).}%
    \thanks{Giampaolo Ferraioli is with the Dipartimento di Scienze e Tecnologie, Università degli Studi di Napoli “Parthenope,” 80143 Naples, Italy, and also with the National Inter-University Consortium for Telecommunications (CNIT), 80126 Naples, Italy (e-mail: giampaolo.ferraioli@uniparthenope.it).}% <-this % stops a space
    }
    
% note the % following the last \IEEEmembership and also \thanks - 
% these prevent an unwanted space from occurring between the last author name
% and the end of the author line. i.e., if you had this:
% 
% \author{....lastname \thanks{...} \thanks{...} }
%                     ^------------^------------^----Do not want these spaces!
%

% The paper headers
\markboth{Journal of \LaTeX\ Class Files,~Vol.~14, No.~8, August~2015}
{Shell \MakeLowercase{\textit{et al.}}: Bare Demo of IEEEtran.cls for IEEE Journals}

% If you want to put a publisher's ID mark on the page you can do it like
% this:
%\IEEEpubid{0000--0000/00\$00.00~\copyright~2015 IEEE}
% Remember, if you use this you must call \IEEEpubidadjcol in the second
% column for its text to clear the IEEEpubid mark.

% use for special paper notices
%\IEEEspecialpapernotice{(Invited Paper)}

% make the title area
\maketitle

% As a general rule, do not put math, special symbols or citations
% in the abstract or keywords.

\begin{abstract}
Speckle is an inherent characteristic of synthetic aperture radar (SAR) imagery that complicates image interpretation and the extraction of reliable information. Although deep-learning-based despeckling methods have demonstrated promising performance, their applicability across different polarimetric configurations is often limited by the need for specialized architectures or separate training procedures. This paper presents a unified deep-learning framework for despeckling single-, dual-, and full-polarization SAR data using a single trained convolutional neural network. The proposed approach adopts a common nine-channel representation of the polarimetric covariance matrix and a polarization-aware masking strategy to accommodate different configurations without modifying the network architecture. 
Qualitative and quantitative comparison on three different scenes from different sensors demonstrate competitive performance across the considered experiments and support the feasibility of a unified, polarization-flexible approach to SAR despeckling across different  polarimetric configurations.
\end{abstract}

% Note that keywords are not normally used for peerreview papers.


% For peer review papers, you can put extra information on the cover
% page as needed:
% \ifCLASSOPTIONpeerreview
% \begin{center} \bfseries EDICS Category: 3-BBND \end{center}
% \fi
%
% For peerreview papers, this IEEEtran command inserts a page break and
% creates the second title. It will be ignored for other modes.
\IEEEpeerreviewmaketitle

\section{Introduction}
\label{sec:introduction}
\textcolor{blue}{THIS MANUSCRIPT IS UNDER REVIEW TO IEEE TGRS.}
\IEEEPARstart{S}{ynthetic} Aperture Radar (SAR) is a fundamental Earth observation technology because of its capability to acquire high-resolution images independently of daylight and weather conditions. These properties make SAR suitable for a wide range of applications, including land-cover mapping, environmental monitoring, agriculture, forestry, disaster assessment, and infrastructure surveillance.
A major limitation of coherent SAR imaging is speckle \cite{Moreira2013}, a signal-dependent fluctuation generated by the coherent combination of multiple elementary scatterers within the same resolution cell. Speckle reduces radiometric accuracy and image interpretability and may negatively affect subsequent processing tasks such as classification, segmentation, change detection, and physical-parameter estimation.
For this reason, a wide variety of despeckling methods has been developed over the last decades  tackling the despeckling issues  with different approaches (local, non-local, deep-learning, etc...) and for different SAR configurations (single-channel intensity, polarimetric and multi-channel). 
Detailed descriptions are provided in the following overviews: \cite{Argenti2013, Moreira2013, Deledalle2014, Ma2018,Fracastoro2021}.

Local filters \cite{Lee1980,Cloude1997,Freeman1998} were long the standard approach to despeckling SAR intensity images, until the non-local paradigm \cite{Parrilli2012,Deledalle2017MuLog,Ferraioli2019} proved more effective at balancing the trade-off between texture retention and edge preservation. Over the last decade, deep learning methods have reached state-of-the-art performance \cite{Vitale2023GMonet,Dalsasso2022}. A similar trend can be observed for polarimetric SAR (PolSAR) data, where methods originally proposed for the single-channel (intensity) case have typically been extended to the polarimetric setting with comparable results \cite{Moreira2013,Ma2018}. However, unlike in single-channel despeckling, deep learning methods for PolSAR despeckling (e.g., \cite{Lu2024Hybrid,Mendes2025PolSAR2PolSAR}), while showing promising performance, have not yet established themselves as the state of the art, with non-local approaches \cite{Deledalle2015NLSAR,Deledalle2017MuLog} still achieving better results. 
% Therefore, in PolSAR imagery, the problem is particularly challenging because the information of interest is represented by complex covariance or coherency matrices and non-local based despeckling method are the ones that best attenuate fluctuations and preserve spatial details, and also maintain the inter-channel correlations and polarimetric properties encoded in these matrices.

% Among the available acquisition modalities, polarimetric SAR (PolSAR) exploits multiple transmit--receive polarization channels to characterize the electromagnetic scattering behavior of natural and man-made targets. The resulting polarimetric measurements provide information about scattering mechanisms that cannot be retrieved from single-channel intensity observations alone \cite{Lee2009,Cloude1997,Freeman1998}.

 % In PolSAR imagery, the problem is particularly challenging because the information of interest is represented by complex covariance or coherency matrices. Therefore, an effective despeckling method must not only attenuate fluctuations and preserve spatial details, but also maintain the inter-channel correlations and polarimetric properties encoded in these matrices \cite{Lee1980,Lee1999PolSAR}.
 In this manuscript, rather than focusing on a specific SAR configuration, the attention is moved on generic multi-channel SAR despeckling. Due to the plethora of SAR systems providing data with different numbers of polarizations, defining a single solution that can handle each of SAR configuration is crucial. The idea is to define a single solution that should take care of preserving spatial textures of the intensity data and, at the same time,  maintain the inter-channel correlations and polarimetric properties accordingly to the data under test. In the last decade, non-local methods such as \cite{Deledalle2015NLSAR,Deledalle2017MuLog,Aghababaei2022MEETSAR} and deep-learning based ones such as \cite{Loic2025} have been proposed as despeckling methods operating in multi-channel configurations. More specifically, NL-SAR \cite{Deledalle2015NLSAR} introduced a unified probabilistic framework for SAR, PolSAR, and InSAR data, combining patch-based similarity with locally selected estimators to obtain a favorable trade-off between noise suppression and resolution preservation. MuLog \cite{Deledalle2017MuLog} employs a matrix-logarithm-based variance-stabilization strategy that enables Gaussian denoisers to be embedded into a multichannel SAR restoration framework. The nonlocal model-free method commonly denoted as MEET-SAR \cite{Aghababaei2022MEETSAR} avoids prescribing a specific speckle distribution and evaluates similarity through empirical patch statistics, making it applicable to both single- and multichannel SAR data . The MuchaPro solution \cite{Loic2025} addresses the problem of filtering data with different dimensionality and provide a deep-learning based approach in which the generic SAR data is decomposed on a set of orthogonal bases and each of them treated as single-channel despeckling.

This work proposes a unified, polarization-flexible deep-learning framework for multi-channel SAR despeckling named \textbf{\prop{}} (\textbf{P}olarization-\textbf{A}ware \textbf{N}etowrk for \textbf{D}especkling \textbf{A}pplication). The method operates directly in the SAR domain, filtering the data through their covariance matrix, and is designed to account for the statistical and physical characteristics of both single-channel and multi-channel SAR data. The SAR data, independently from the single or multi-polarimetric configuration, is represented by a nine-channel real-valued  array derived from the polarimetric covariance matrix, allowing observations with different polarimetric dimensionalities to be expressed within the same input space.
The training dataset is based on real full-polarimetric UAVSAR observations and hybrid approach \cite{Vitale2022} is used for the derivation of the ground truth. The architecture used in \prop{} is a single convolutional neural network (CNN) designed and trained to multi-polarimetric  SAR configurations rather than considering a separate training for a specific configuration (single, dual, full polarization). A polarization-aware masking strategy identifies the available covariance components and enables the CNN to process different polarimetric configurations without architectural modifications or configuration-specific retraining.
 % During training, polarization-aware masking is employed to expose the network to single-, dual-, and full-polarimetric input configurations.
% acquired over the same geographical area at different acquisition times. Since noise-free SAR measurements are not available, low-speckle pseudo-reference covariance matrices are generated from 27 co-registered acquisitions through multitemporal averaging followed by spatial filtering. 
The proposed solution is then tested on three different sensors over three testing scenarios independent from the training data. For each sensor, single- and dual-polarization configurations are obtained by selecting the corresponding covariance components from the available full-polarization observations. The proposed method is compared, qualitatively and quantitatively, with multi-channel despeckling methods and with specific deep-learning baseline for specific channel configuration whose codes are available online.
% The evaluation considers speckle suppression, radiometric consistency, and the statistical properties of the despeckling residuals, including covariance-trace and maximum-eigenvalue distributions.

The main contributions of this work are summarized as follows:
\begin{itemize}

    \item A unified polarization-flexible deep-learning framework for SAR despeckling operating directly in the SAR domain taking care at once characteristics and properties of single and multi-channel configuration.

    \item A channel-masking strategy that makes a single neural network aware  and generic to the input dimensionality, allowing it to process inputs with a variable number of channels without architectural changes.
\end{itemize}

The remainder of this paper is organized as follows. Section \ref{sec:method} introduces the proposed unified framework. Section \ref{sec:experimental_setup} describes the experimental setup and evaluation methodology. Section \ref{sec:results} presents the experimental results. Section \ref{sec:discussion} discusses the main findings and limitations of the proposed approach. Finally, Section \ref{sec:conclusion} presents the conclusions and describes possible future developments.

\section{Proposed Methodology}
\label{sec:method}

This section presents the proposed unified framework for the despeckling of single-, dual-, and fully polarimetric SAR data. The main objective is to process covariance matrices of different polarimetric dimensionalities using a single neural network, without modifying its architecture or retraining it for each polarimetric configuration. To this end, the input data are mapped onto a common real-valued representation, while the unavailable polarimetric components are explicitly excluded from the restoration process.

\subsection{Problem Formulation}
\label{subsec:problem_formulation}

Let $\mathbf{k}\in\mathbb{C}^{p}$ denote the complex scattering vector associated with a given polarimetric acquisition, where $p$ depends on the available polarization channels. Under the multilook complex Wishart model, the observed covariance matrix can be expressed as

\begin{equation}
    \mathbf{C}
    =
    \frac{1}{L}
    \sum_{\ell=1}^{L}
    \mathbf{k}_{\ell}\mathbf{k}_{\ell}^{H},
    \label{eq:sample_covariance}
\end{equation}

where $L$ is the number of looks and $(\cdot)^H$ denotes the Hermitian transpose. The matrix $\mathbf{C}\in\mathbb{C}^{p\times p}$ is Hermitian and positive semidefinite, and contains both the backscattered power of the individual channels and their complex cross-correlations \cite{Deledalle2017MuLog}.

Depending on the acquisition mode, the covariance matrix assumes different dimensions. In the single-polarimetric case, the observation reduces to a scalar intensity,

\begin{equation}
    \mathbf{C}_{\mathrm{SP}} =
    \begin{bmatrix}
        C_{11}
    \end{bmatrix}.
\end{equation}

For a dual-polarimetric acquisition, the observation is represented by

\begin{equation}
    \mathbf{C}_{\mathrm{DP}} =
    \begin{bmatrix}
        C_{11} & C_{12} \\
        C_{12}^{*} & C_{22}
    \end{bmatrix},
\end{equation}

whereas fully polarimetric data are described by the $3\times3$ covariance matrix

\begin{equation}
    \mathbf{C}_{\mathrm{FP}} =
    \begin{bmatrix}
        C_{11} & C_{12} & C_{13} \\
        C_{12}^{*} & C_{22} & C_{23} \\
        C_{13}^{*} & C_{23}^{*} & C_{33}
    \end{bmatrix}.
    \label{eq:full_covariance}
\end{equation}

The despeckling problem consists of estimating the underlying noise-free covariance matrix $\boldsymbol{\Sigma}$ from the noisy observation $\mathbf{C}$ while preserving spatial details, radiometric information, and the relationships between the polarimetric channels. The proposed estimator is written as

\begin{equation}
    \widehat{\boldsymbol{\Sigma}}
    =
    f_{\boldsymbol{\theta}}
    \left(
        \mathbf{C},\mathbf{M}
    \right),
    \label{eq:unified_estimator}
\end{equation}

where $f_{\boldsymbol{\theta}}$ denotes the proposed neural network, parameterized by $\boldsymbol{\theta}$, and $\mathbf{M}$ identifies the components available in the considered polarimetric configuration.

\subsection{Training Dataset Generation}
\label{subsec:training_dataset}

The first issue to be tackled for designing a deep learning solution is the construction of reliable training dataset. Obtaining reliable ground-truth data for SAR despeckling is inherently difficult due to the lack of a noise-free observation of the same scene. Therefore, the network was trained with a self-supervised approach according to which the noise-free reference covariance matrices were generated from a multitemporal SAR stack of real fully polarimetric acquisitions following the hybrid approach \cite{Vitale2022}.

The noise-free reference estimation relies on $T$ fully polarimetric acquisitions collected over the same geographical area at different acquisition times. The acquisitions are spatially co-registered to ensure pixel-wise correspondence across the temporal stack. For each acquisition $t$, the reciprocal lexicographic scattering vector is defined as

\begin{equation}
    \mathbf{k}^{(t)}
    =
    \begin{bmatrix}
        S_{HH}^{(t)} &
        \sqrt{2}S_{HV}^{(t)} &
        S_{VV}^{(t)}
    \end{bmatrix}^{T},
    \label{eq:training_scattering_vector}
\end{equation}

and the corresponding sample covariance matrix     $\mathbf{C}^{(t)}$ is obtained as in Equation \ref{eq:sample_covariance}

% \begin{equation}
%     \mathbf{C}^{(t)}
%     =
%     \mathbf{k}^{(t)}
%     \mathbf{k}^{(t)H}.
%     \label{eq:single_acquisition_covariance}
% \end{equation}

The resulting covariance matrices are subsequently averaged along the temporal dimension to obtain a multitemporal covariance estimate.

\begin{equation}
    \overline{\mathbf{C}}_{\mathrm{MT}}
    =
    \frac{1}{T}
    \sum_{t=1}^{T}
    \mathbf{C}^{(t)},
    % \qquad T=27.
    \label{eq:multitemporal_covariance}
\end{equation}
       
Multitemporal averaging substantially reduces speckle because the acquisitions provide different realizations of the scattering process. The noise-free covariance matrix $\mathbf{\Sigma}_{ref}$ is then obtained as: 
\begin{equation}
    \mathbf{\Sigma}_{ref} = w(\overline{\mathbf{C}}_{\mathrm{MT}})
\label{eq: spatial filtering}
\end{equation}

where $w(\cdot)$ is a spatial filter to reduce residual spatial fluctuations.

This procedure provides a good estimation of noise-free reference if the multi-temporal stacks is quite stationary over time. Following the hybrid approach, the noisy input $\mathbf{C_{in}}$ is then obtained by adding Wishart distributed noise (fully-developed hypothesis) to the $\mathbf{\Sigma}_{ref}$ as:

\begin{equation}
    \mathbf{C_{in}} = \mathbf{\Sigma}_{ref}^{1/2} \mathbf{S} \mathbf{\Sigma}_{ref}^{1/2}
\end{equation}

where $\mathbf{S} \sim W(\mathbf{\Sigma}_{ref},L)$ models the speckle fluctuations for multivariate SAR images \cite{Deledalle2017MuLog}.

This derivation of the input noisy data avoid to embed in the network learning a wrong relationship between input and reference data overcoming the issue related to temporal changes.

\subsection{Unified Polarimetric Representation}
\label{subsec:unified_representation}

Directly processing covariance matrices of different sizes would normally require configuration-specific network architectures. Instead, the proposed framework embeds every observation into a common representation derived from the fully polarimetric $3\times3$ covariance matrix.

Each noisy/reference pair $(\mathbf{C_{in}}, \mathbf{\Sigma}_{ref})$ are constructed from fully polarimetric data and the single-, dual-, and fully polarimetric samples are obtained by masking the not available polarimetric channels. Therefore, all polarimetric configurations share the same spatial content and input-reference generation procedure, allowing the network to learn from the three polarimetric dimensionalities within a common training framework.

Since a covariance matrix is Hermitian, its independent information can be represented by the upper triangular matrix that can be shaped in nine real-valued components. The proposed representation first collects the three real-valued diagonal elements, followed by the real parts of the three upper-triangular cross-covariance terms and, finally, their imaginary parts. Accordingly, the fully polarimetric covariance matrix is mapped onto the nine-channel real-valued tensor

\begin{equation}
\begin{split}
    \mathcal{R}\left(\mathbf{C}_{in}\right)
    =
    [&C_{11}, C_{22}, C_{33},\\
    &\operatorname{Re}(C_{12}),
     \operatorname{Re}(C_{13}),
     \operatorname{Re}(C_{23}),\\
    &\operatorname{Im}(C_{12}),
     \operatorname{Im}(C_{13}),
     \operatorname{Im}(C_{23})].
\end{split}
\label{eq:real_representation}
\end{equation}

Single- and dual-polarimetric observations are embedded into the same nine-channel space. Components associated with unavailable polarimetric measurements are assigned a neutral value and identified through a binary availability mask,

\begin{equation}
    \mathbf{M}
    =
    \left[m_1,m_2,\ldots,m_9\right],
    \qquad
    m_i\in\{0,1\},
    \label{eq:availability_mask}
\end{equation}

where $m_i=1$ indicates an observed covariance component and $m_i=0$ indicates an unavailable component. The effective network input is then obtained as

\begin{equation}
    \mathbf{X}
    =
    \mathbf{M}\odot\mathcal{R}(\mathbf{C}_{in}),
    \label{eq:masked_input}
\end{equation}

where $\odot$ denotes element-wise multiplication, with the mask broadcast over the spatial dimensions.

The same representation is applied to the network output,

\begin{equation}
    \widehat{\mathbf{Y}}
    =
    \mathbf{M}\odot
    f_{\boldsymbol{\theta}}(\mathbf{X}),
    \label{eq:masked_output}
\end{equation}

so that the reconstruction is restricted to physically available polarimetric components. The resulting channels are subsequently mapped back to their Hermitian covariance form through the inverse transformation

\begin{equation}
    \widehat{\boldsymbol{\Sigma}}
=\mathcal{R}^{-1}(\mathbf{\widehat{Y}}) =
\begin{bmatrix}
\widehat{y}_1 & \widehat{y}_4+\mathrm{j}\widehat{y}_7 & \widehat{y}_5+\mathrm{j}\widehat{y}_8 \\
\widehat{y}_4-\mathrm{j}\widehat{y}_7 & \widehat{y}_2 & \widehat{y}_6+\mathrm{j}\widehat{y}_9 \\
\widehat{y}_5-\mathrm{j}\widehat{y}_8 & \widehat{y}_6-\mathrm{j}\widehat{y}_9 & \widehat{y}_3
\end{bmatrix},
\label{eq:inverse_real_representation}
\end{equation}

The inverse transformation reconstructs a Hermitian covariance matrix from the nine-channel network output. Although Hermitian symmetry is preserved by construction, positive definiteness is not explicitly enforced by the network.

This formulation provides a fixed input and output dimensionality for all the considered acquisition modes. Consequently, the same set of learned parameters can be used for single-, dual-, and fully polarimetric data, without configuration-specific architectural modifications.
\subsection{Unified Training Strategy}
\label{subsec:training_strategy}
A key aspect of the proposed training strategy is that all polarimetric configurations are derived from the same fully polarimetric training data. Specifically, both the noisy full-polarimetric covariance matrix $\mathbf{C}$ and its reference $\boldsymbol{\Sigma}$ are first mapped onto the nine-channel representation introduced in Section \ref{subsec:unified_representation}. Single-, dual-, and fully polarimetric training samples are then generated by applying the corresponding availability masks. 

For a generic configuration $q \in \{\mathrm{SP},\mathrm{DP},\mathrm{FP}\}$, the resulting training pair is

\begin{equation}
    \mathbf{X}_{q}
    =
    \mathbf{M}_{q} \odot \mathcal{R}(\mathbf{C}),
    \qquad
    \mathbf{Y}_{q}
    =
    \mathbf{M}_{q} \odot \mathcal{R}(\boldsymbol{\Sigma}),
\end{equation}

where $\mathbf{M}_{q}$ identifies the covariance components available for configuration $q$. Thus, the different polarimetric configurations share the same underlying scene content and reference data and differ only in the subset of covariance components made available to the network.

\begin{figure*}[!t]
    \centering
    \includegraphics[width=0.8\paperwidth]
    {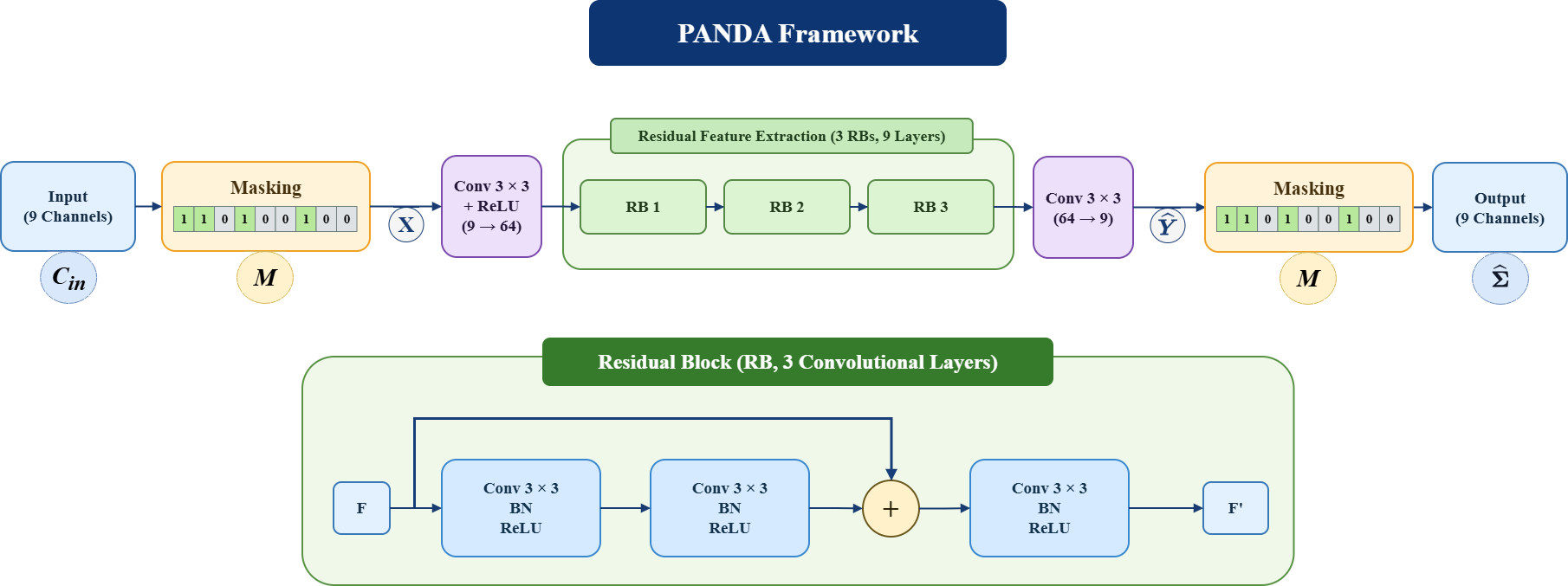}
    \caption{Architecture of the adopted 11-layer convolutional neural
    network, adapted from MONet \cite{Vitale2021MONet}. The first and final
    layers are modified to process the proposed nine-channel covariance
    representation, while the internal feature-extraction structure is
    retained.}
    \label{fig:network_architecture}
\end{figure*}

Samples corresponding to the different masks are jointly used during training to optimize a single set of network parameters. This exposes the model to different levels of available polarimetric information while maintaining a fixed input and output dimensionality. The network is therefore encouraged to learn a shared despeckling function rather than independent mappings specialized for each polarimetric configuration.

At inference time, the mask corresponding to the considered acquisition configuration is applied to the input using the same representation adopted during training. The network architecture and learned parameters remain unchanged regardless of the available polarimetric dimensionality, and no configuration-specific retraining or fine-tuning is performed on the test sensors.

A single model is trained using this strategy, and the resulting checkpoint is used unchanged for all subsequent experiments. The same network parameters are therefore employed for all polarization configurations and test sensors, without configuration-specific retraining or sensor-specific fine-tuning.

\subsection{Network Architecture}
\label{subsec:network_architecture}

The proposed architecture is a straightforward fully convolutional neural network with skip connections every two layers. It comprises 11 convolutional layers. The first layer maps the nine input channels onto 64 feature maps. The following nine internal layers operate on 64 feature maps using $3\times3$ convolutional kernels, batch normalization, ReLU activations, and residual connections. The overall architecture is shown in Fig.~\ref{fig:network_architecture}.
The choice of this architecture follows the ResNet \cite{he2016} rationale of using skip connections as simple way to propagate low level features and contrasting the vanishing gradient issue. 

The network accepts nine-channel input $\mathbf{X}$, no spatial downsampling or upsampling is employed, so the original image resolution is preserved throughout the network. The final convolutional layer maps the extracted features back to nine output channels and does not use batch normalization or an activation function, allowing both positive and negative values to be estimated for the real and imaginary cross-covariance components.

The availability mask $\mathbf{M}$ is applied to the input and as well as to the output to retain only the components associated with the considered acquisition mode. The resulting vector is then converted back into its Hermitian matrix form through the inverse transformation $\mathcal{R}^{-1}(\cdot)$. The same network parameters are used for single-, dual-, and fully polarimetric data; therefore, no configuration-specific architectural changes or retraining are required.

\subsection{Loss Function}
\label{subsec:loss_function}

The network is optimized using a loss function designed to jointly promote covariance reconstruction accuracy and spatial-detail preservation. It is defined as

\begin{equation}
    \mathcal{L}
    =
    \mathcal{L}_{\mathrm{frob}}
    +
    \mathcal{L}_{\mathrm{grad}}.
    \label{eq:total_loss}
\end{equation}

The first term measures the normalized reconstruction error between the estimated output $\widehat{\mathbf{Y}}$ and the corresponding reference $\mathbf{Y}$:

\begin{equation}
    \mathcal{L}_{\mathrm{frob}}
    =
    \mathbb{E}
    \left[
    \frac{
        \left\|
        \widehat{\mathbf{Y}}-\mathbf{Y}
        \right\|_{F}
    }{
        \left\|\widehat{\mathbf{Y}}\right\|_{F}
        +
        \left\|\mathbf{Y}\right\|_{F}
        +\epsilon
    }
    \right],
    \label{eq:frobenius_loss}
\end{equation}

where $\|\cdot\|_{F}$ denotes the Frobenius norm and $\epsilon$ is a small constant introduced for numerical stability. The normalization reduces the dependence of the loss on the absolute backscattering intensity.

To improve the preservation of edges and spatial structures, a second term compares the gradients of the estimated and reference data. The gradient representation is obtained from the horizontal and vertical finite differences as

\begin{equation}
    d_{\mathrm{gr}}(\mathbf{I})
    =
    \left|d_x(\mathbf{I})\right|
    +
    \left|d_y(\mathbf{I})\right|.
    \label{eq:gradient_operator}
\end{equation}

The corresponding loss is

\begin{equation}
    \mathcal{L}_{\mathrm{grad}}
    =
    \mathbb{E}
    \left[
    \frac{
        \left|
        d_{\mathrm{gr}}(\widehat{\mathbf{Y}})
        -
        d_{\mathrm{gr}}(\mathbf{Y})
        \right|
    }{
        \left|d_{\mathrm{gr}}(\widehat{\mathbf{Y}})\right|
        +
        \left|d_{\mathrm{gr}}(\mathbf{Y})\right|
        +\epsilon
    }
    \right].
    \label{eq:gradient_loss}
\end{equation}

Both loss terms are computed using the masked network output and the corresponding masked pseudo-reference, as defined in Section III-D. Consequently, covariance components that are unavailable in the considered polarimetric configuration are excluded from both the reconstruction and gradient losses. The expectation is evaluated over the training samples, spatial positions, and available covariance channels.

\section{Experimental Setup}
\label{sec:experimental_setup}

This section describes the datasets employed for training and testing together with implementation details of the solution, the competing despeckling methods, and the evaluation protocol adopted for the single-, dual-, and fully polarimetric configurations.

\subsection{Implementation Details}
\label{subsec:implementation_details}
The training dataset construction relies $T=27$ fully polarimetric UAVSAR acquisitions collected over the same geographical area at different acquisition times.  MuLog \cite{Deledalle2017MuLog} has been applied as spatial filtering process $w(\cdot)$ in Equation \ref{eq: spatial filtering}.
The data is split into patches of $64\times64$ pixels and the resulting dataset contains $100\,602$ training patches and $11\,178$ validation patches.

The network is trained for 120 epochs using the Adam optimizer, with a batch size of 128 and an initial learning rate of $10^{-3}$. The learning rate is reduced by a factor of 0.5 every 30 epochs.

The model checkpoint achieving the lowest validation loss is selected for the subsequent experiments. The same checkpoint is used for all test sensors and polarimetric configurations, without additional training or sensor-specific fine-tuning.

\subsection{Test Datasets and Polarimetric Configurations}
\label{subsec:test_datasets}

Three testing cases have been considered for assessing the performance of \prop{}. The three testing cases, listed in Table~\ref{tab:test_sensors}, are three fully polarimetric images acquired over different areas and from different sensors operating at different frequencies. ALOS-PALSAR operates at L-band  and the area over the bay of San Francisco in California (USA) including urban, vegetation, and sea is selected; RADARSAT-2 operates at C band and the rural area over Flevoland in The Netherlands is selected; BIOMASS operates in the P band and the area towards the Paraguayan coast of lake Paranà is selected.

\begin{table}[h]
    \centering
    \caption{SAR sensors employed for the cross-sensor evaluation.}
    \label{tab:test_sensors}
    \renewcommand{\arraystretch}{1.15}
    \begin{tabular}{lccc}
        \hline
        \textbf{Sensor} & \textbf{Band} & \textbf{Center frequency} & \textbf{Area}\\
        \hline
        ALOS PALSAR & L & 1.270 GHz  & San Francisco (USA)\\
        RADARSAT-2  & C & 5.405 GHz & Flevoland (NL)\\
        BIOMASS     & P & 0.435 GHz & Paranà Lake (PY)\\
        \hline
    \end{tabular}
\end{table}

% ALOS PALSAR provides L-band observations, whose relatively long wavelength makes them particularly suitable for the analysis of vegetation and distributed natural targets. RADARSAT-2 operates at C-band and provides a substantially different acquisition domain in terms of wavelength, resolution, and sensor characteristics. Finally, BIOMASS provides fully polarimetric P-band observations characterized by a considerably longer wavelength and stronger penetration into vegetation volumes.
The three testing datasets have been selected to evaluate of the ability of \prop{} to generalize to different scenarios and beyond the L-band UAVSAR data used during training.

Starting from each fully polarimetric acquisition, the covariance matrices are converted into the common nine-channel representation. Single- and dual-polarimetric configurations are then emulated by applying the corresponding availability masks described in Section~\ref{subsec:unified_representation}.

This procedure ensures that the three configurations associated with each sensor represent exactly the same spatial content. Consequently, variations among the results can be attributed to the available polarimetric information rather than to differences in the observed scene. The same trained network and model parameters are used in all cases, without configuration-specific fine-tuning.

\subsection{Compared Despeckling Methods}
\label{subsec:compared_methods}
For a fair comparison, methods whose code are available on-line have been selected. The comparison is performed on all the three possible polarimetric configuration (single-, dual- and full) for each test dataset. The proposed method is compared with two established non-local filers, namely MuLog \cite{Deledalle2017MuLog} and MEET-SAR \cite{Aghababaei2022MEETSAR}. Both methods support covariance data with different polarimetric dimensionalities and are therefore applied to all the considered configurations.

MuLog transforms the multichannel multiplicative-speckle problem into an approximately additive Gaussian denoising problem through a matrix logarithmic transformation. This formulation enables the application of a Gaussian denoiser within an iterative restoration framework. In the experiments, MuLog is applied independently to the single-, dual-, and fully polarimetric covariance representations.

MEET-SAR is a nonlocal model-free method specifically designed for both single- and multichannel SAR data. It exploits the similarity between spatial patches to suppress speckle while preserving radiometric and interchannel information. Unlike MuLog, it does not require an explicit parametric model of the underlying reflectivity.

In addition, a well assessed deep learning based method specifically designed for the polarimetric configuration under test is selected for each acquisition mode. Specifically, CL-SAR \cite{Fang2024CLSAR} is considered for the single-polarization case, CNN-PolSAR \cite{MestreQuereda2025} for the dual-polarization case, and
DRUNet \cite{Zhang2022DRUNet} for the fully polarimetric case. 
A scheme of the testing framework is depicted in Table \ref{tab:compared_methods}.

In order to not introduce bias, all methods are tested using original shared codes, applied to the same input covariance matrices and evaluated over the same spatial regions. It is worthy to remind that no configuration-specific fine-tuning of \prop{} is performed on the test datasets.

\begin{table}[h]
\centering
\caption{Applicability of the compared despeckling methods to the considered polarimetric configurations.}
\label{tab:compared_methods}
\renewcommand{\arraystretch}{1.15}
\begin{tabular}{lccc}
\hline
\textbf{Method} & \textbf{Single-pol} & \textbf{Dual-pol} & \textbf{Full-pol} \\
\hline
MuLog \cite{Deledalle2017MuLog}
    & $\checkmark$ & $\checkmark$ & $\checkmark$ \\
MEET-SAR \cite{Aghababaei2022MEETSAR}
    & $\checkmark$ & $\checkmark$ & $\checkmark$ \\
CL-SAR \cite{Fang2024CLSAR}
    & $\checkmark$ & -- & -- \\
CNN-PolSAR \cite{MestreQuereda2025}
    & -- & $\checkmark$ & -- \\
DRUNet \cite{Zhang2022DRUNet}
    & -- & -- & $\checkmark$ \\
\prop{}
    & $\checkmark$ & $\checkmark$ & $\checkmark$ \\
\hline
\end{tabular}
\end{table}

\subsection{Evaluation Metrics}
\label{subsec:evaluation_metrics}

Since noise-free references are unavailable for the considered real SAR acquisitions, the methods are evaluated using complementary no-reference criteria. The assessment considers speckle suppression, preservation of radiometric and spatial information, and consistency with the expected polarimetric speckle statistics. Due the different nature of single-channel data and multi-polarimetric data, two different quantitative evaluation metric sets have been considered:

\subsubsection{Single-Channel Metrics}

%Speckle reduction in homogeneous regions is measured through the equivalent number of looks (ENL) \cite{Anfinsen2009ENL}, computed as

%\begin{equation}
%\mathrm{ENL} = \frac{\mu^{2}}{\sigma^{2}},
%\end{equation}

%where $\mu$ and $\sigma^{2}$ are the sample mean and variance within a homogeneous region, respectively. Higher ENL values indicate stronger speckle suppression. However, since high ENL values may also result from excessive smoothing, this metric is not considered independently.
For the single-channel case, the M-index \cite{Gomez2017MIndex}, Mean of Ratio (MoR), the Variance of Ratio (VoR), and the Mean of Image (MoI)\cite{11028079} have been carried out.
\begin{itemize}

\item \textbf{M-Index.} The no-reference M-index is employed to assess the statistical and structural properties of the ratio image. It accounts for deviations of its mean and ENL \cite{Anfinsen2009ENL} from the expected speckle statistics, together with the amount of spatial structure remaining in the residual. Lower M-index values indicate a better balance between speckle suppression and information preservation. For consistency across the different polarimetric configurations, the M-index is computed from the total backscattered power, represented by the covariance-matrix trace.

    \item \textbf{MoR.} Measures the mean of the ratio image $R = \mathbf{C}_{in}/\widehat{\Sigma}$, where $\mathbf{C}_{in}$ and $\widehat{\Sigma}$ denote the noisy and filtered images, respectively. For an ideal despeckling filter, the ratio should preserve the unit mean of the multiplicative speckle noise, yielding $\mathrm{MoR} = 1$.

    \item \textbf{VoR.} Measures the variance of the ratio image $R = \mathbf{C}_{in}/\widehat{\Sigma}$. Under a unit-mean, unit-variance exponential speckle model, an ideal filter is expected to yield $\mathrm{VoR} = 1$.

    \item \textbf{MoI.} Evaluates the preservation of the mean image intensity after filtering. Since multiplicative speckle noise has unit mean, an ideal despeckling filter should preserve the mean intensity of the original noisy image. MoI quantifies the absolute difference between the mean intensities of the filtered and noisy images, with an ideal value of $\mathrm{MoI} = 0$.
\end{itemize}

\subsubsection{Multi-Polarimetric Metrics}
For the multi-polarimetric case, statistical consistency between co-polarized elements and cros-polarized elements is assessed directly in the covariance domain. Given the observed covariance matrix $\mathbf{C}$ and the filtered estimate $\widehat{\mathbf{\Sigma}}$, the latter is corrected using the nearestSPD procedure to obtain a Hermitian positive-definite matrix $\widehat{\mathbf{\Sigma}}_{\mathrm{SPD}}$. This correction is applied consistently to the outputs of all compared despeckling methods, exclusively for statistical evaluation, and is not part of the proposed network architecture or inference procedure. The Cholesky factorization of the corrected estimate is then computed as $\widehat{\mathbf{\Sigma}}_{\mathrm{SPD}} = \mathbf{R}^{H}\mathbf{R}$, where $\mathbf{R}$ is upper triangular. The normalized residual covariance matrix is subsequently obtained as

\begin{equation}
\mathbf{S}
=
\mathbf{R}^{-H}
\mathbf{C}
\mathbf{R}^{-1},
\end{equation}

The empirical probability density functions of the normalized trace $\mathrm{Tr}(\mathbf{S})/p$  and of the largest eigenvalue $\lambda_{\max}(\mathbf{S})$  are compared with reference distributions generated under the complex Wishart model.

Specifically, the reference distributions are approximated through Monte Carlo simulation using $10^6$ independent realizations of complex Gaussian scattering vectors with identity covariance. A single-look model ($L=1$) is adopted for all experiments, while the polarimetric dimension $p$ is set according to the considered acquisition configuration ($p=2$ for dual-polarization and $p=3$ for full-polarization data). This ensures a consistent statistical reference across the three sensors and the corresponding polarimetric configurations.

For an ideal despeckling estimator, the normalized residual covariance matrix is expected to follow the corresponding complex Wishart distribution. Deviations between the empirical and reference distributions therefore provide an indication of residual statistical inconsistencies introduced by the filtering process.

The agreement between the empirical and reference distributions is quantified through the Kullback--Leibler divergence and the PDF overlap coefficient \cite{9006944}:

\begin{equation}
\begin{split}
D_{\mathrm{KL}} &=
\int f_{\mathrm{emp}}(x)
\log\frac{f_{\mathrm{emp}}(x)}{f_{\mathrm{ref}}(x)}\,dx,\\
\mathrm{OVL} &=
\int \min\!\left[
f_{\mathrm{emp}}(x),f_{\mathrm{ref}}(x)
\right]\,dx .
\end{split}
\end{equation}

Accordingly, lower $D_{\mathrm{KL}}$ and higher OVL values indicate better agreement with the expected speckle statistics. All methods are evaluated using the same homogeneous regions, PDF estimator, and sampling grid. Finally, quantitative assessment is complemented by visual inspection of the filtered images and residual maps to identify blurring, edge degradation, or residual scene structures.

\section{Experimental Results}
\label{sec:results}
\begin{figure*}[t]
    \centering
    \begingroup
    \newsavebox{\fullfirstpanel}
    \newcommand{\fulllabelledpanel}[2]{%
        \sbox{\fullfirstpanel}{%
            \includegraphics[width=0.15\textwidth]{#2}}%
        \makebox[0pt][r]{%
            \smash{\raisebox{\dimexpr\ht\fullfirstpanel/2\relax}{%
                \raisebox{-0.5\height}{%
                    \rotatebox{90}{\small\textbf{#1}}}}}%
            \hspace{4pt}}%
        \usebox{\fullfirstpanel}}

    \begin{tabular}{ccccc}
        SAR & MEET-SAT & MuLog & DRUNet & \prop{} \\

        \fulllabelledpanel{RADARSAT-2}{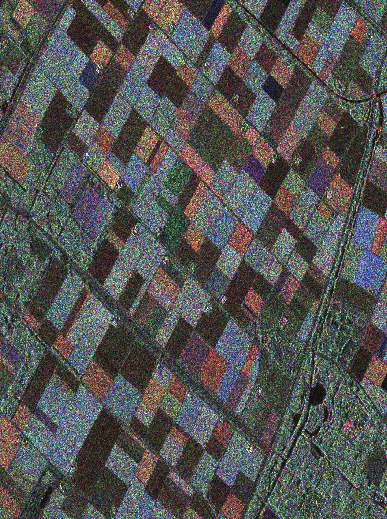}&
        \includegraphics[width=0.15\textwidth]{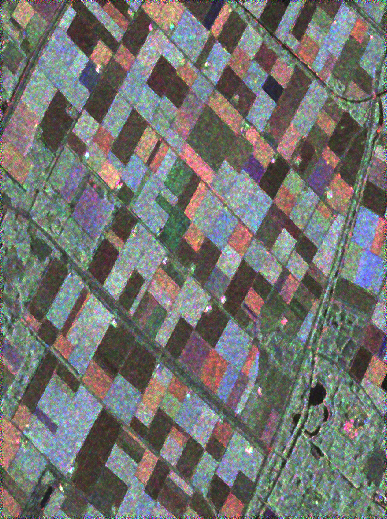}&
        \includegraphics[width=0.15\textwidth]{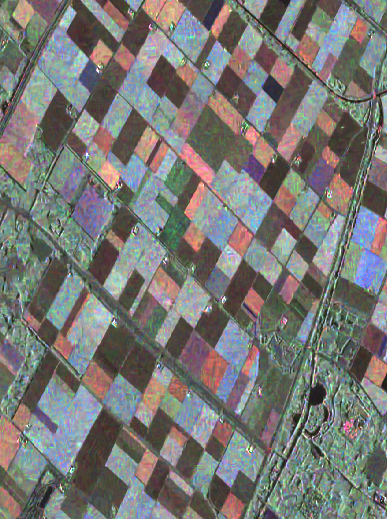}&
        \includegraphics[width=0.15\textwidth]{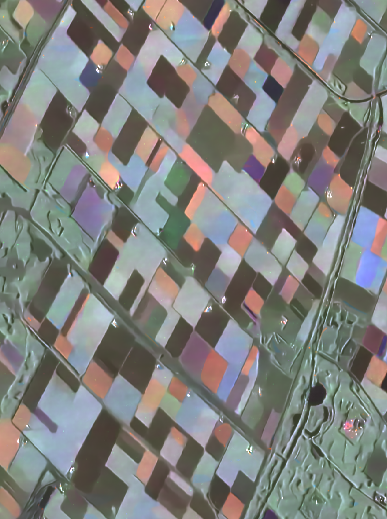}&
        \includegraphics[width=0.15\textwidth]{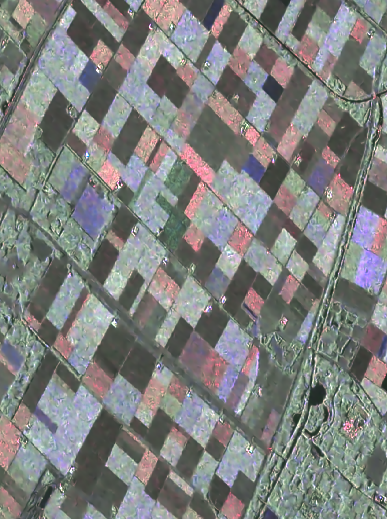}\\

        \fulllabelledpanel{BIOMASS}{Images/FullPol/BIOMASS_fullpol_noisy.png}&
        \includegraphics[width=0.15\textwidth]{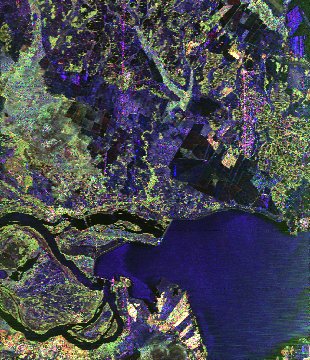}&
        \includegraphics[width=0.15\textwidth]{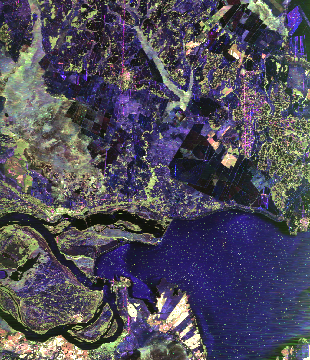}&
        \includegraphics[width=0.15\textwidth]{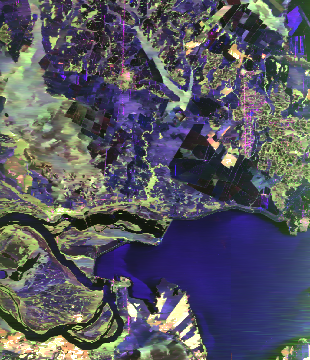}&
        \includegraphics[width=0.15\textwidth]{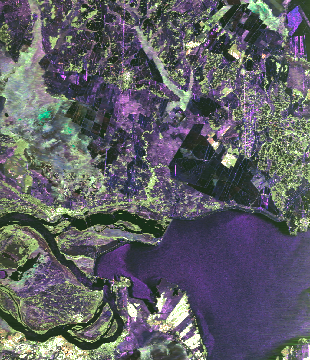}\\

        \fulllabelledpanel{ALOS}{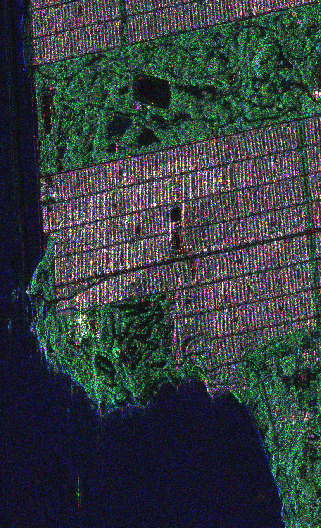}&
        \includegraphics[width=0.15\textwidth]{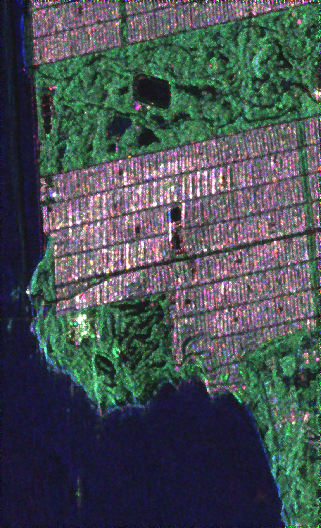}&
        \includegraphics[width=0.15\textwidth]{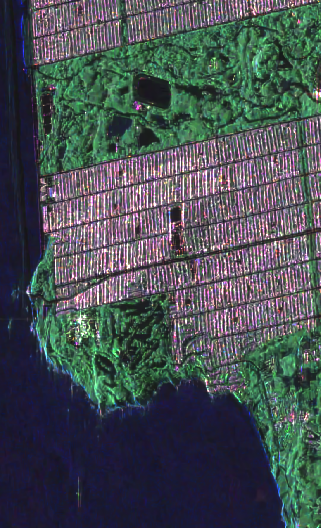}&
        \includegraphics[width=0.15\textwidth]{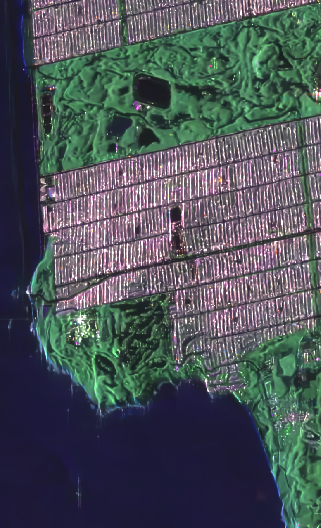}&
        \includegraphics[width=0.15\textwidth]{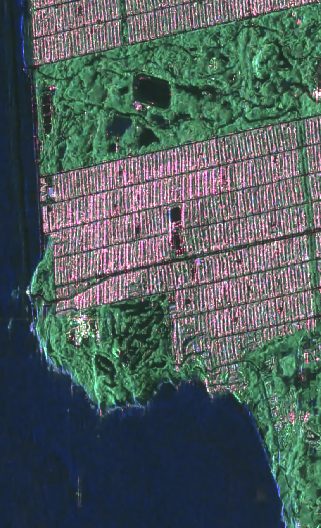}\\
    \end{tabular}
    \endgroup

    \caption{Full-polarization results. From left to right the noisy image and filtering results. From top to bottom RADARSAT-2, BIOMASS and ALOS testing cases. The Pauli composition of polarimetric channels are displayed.}
    \label{fig:full_comparison}
\end{figure*}

In this Section, qualitative and quantitative comparison are reported. The experimental results are organized according to the polarimetric configuration rather than the acquisition sensor. This organization highlights the ability of the proposed model to operate under different amounts of polarimetric information while retaining the same architecture and trained parameters. For each configuration, the proposed method is compared with MuLog, MEET-SAR, and the corresponding configuration-dependent deep-learning baseline. 

\subsection{Full-Polarization Results}

For the full-polarization experiments, \prop{} is compared with the model-based approaches MEET-SAR and MuLog and with DRUNet as the deep-learning baseline.

The qualitative comparison over the three test datasets is shown in Fig.~\ref{fig:full_comparison}: the figure shows from top to bottom the results over RADARSAT-2, BIOMASS and ALOS PALSAR; from left to right the Pauli representation of noisy fully polarimetric data, MEET-SAR, MuLog, DRUNet and PANDA. Generally, MEET-SAR tends to preserve polarimetric information but at the cost of a limited speckle suppression, MuLog looks the best in preserving urban areas and shows a good trade-off between speckle suppression and polarimetric and spatial details preservation but producing some artifacts in vegetated areas, DRUNet shows strong smoothing effect, producing clear homogeneous areas but at the cost of limited polarimetric characteristic preservation. \prop{} achieves a speckle suppression comparable to that of DRUNet, but without introducing over-smoothing, thus better preserving the polarimetric information. At the same time, it preserves spatial details better than MuLog, although at the cost of a slightly lower polarimetric fidelity.

\begin{figure}[h]
     \centering
     \begin{tabular}{ccccc}
     SAR & MEET-SAR & MuLog & DRUNet & \prop{}\\
     \includegraphics[width=0.15\columnwidth]{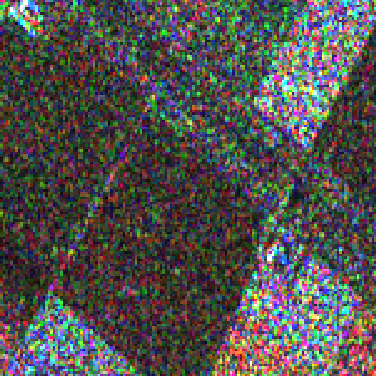}&
     \includegraphics[width=0.15\columnwidth]{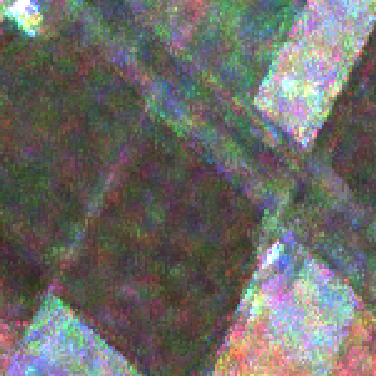}&
     \includegraphics[width=0.15\columnwidth]{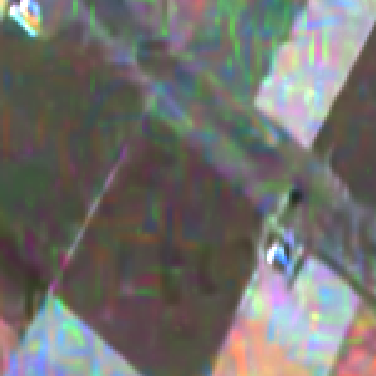}&
     \includegraphics[width=0.15\columnwidth]{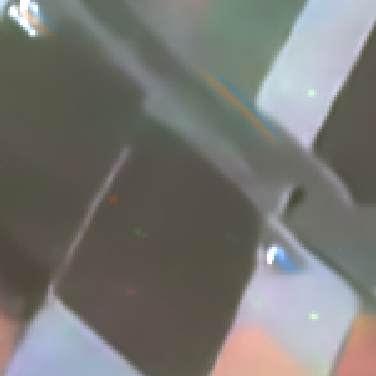}&
     \includegraphics[width=0.15\columnwidth]{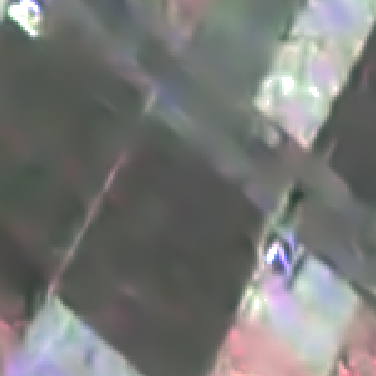}\\
    
     \includegraphics[width=0.15\columnwidth]{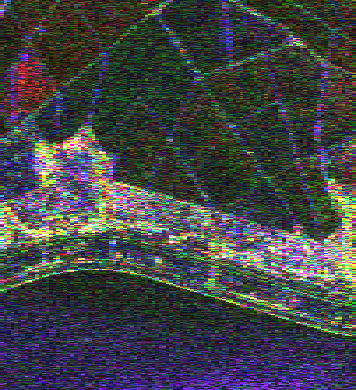}&
     \includegraphics[width=0.15\columnwidth]{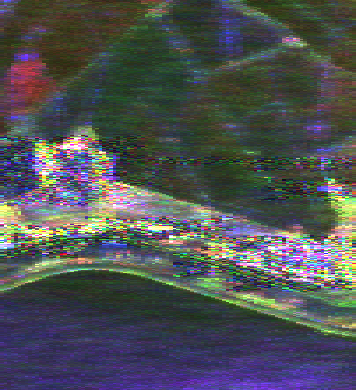}&
     \includegraphics[width=0.15\columnwidth]{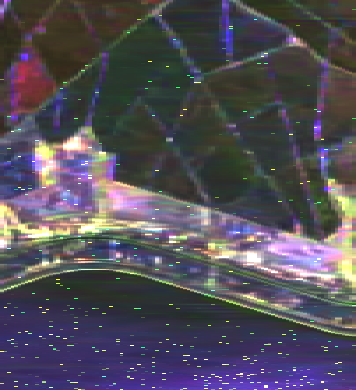}&
     \includegraphics[width=0.15\columnwidth]{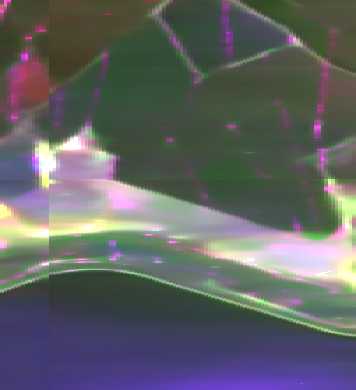}&
     \includegraphics[width=0.15\columnwidth]{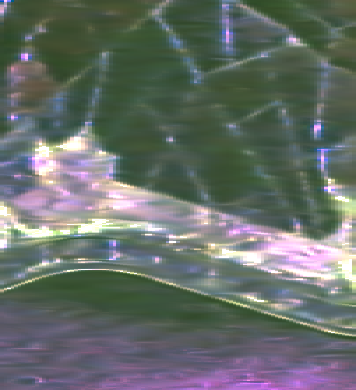}\\

     \includegraphics[width=0.15\columnwidth]{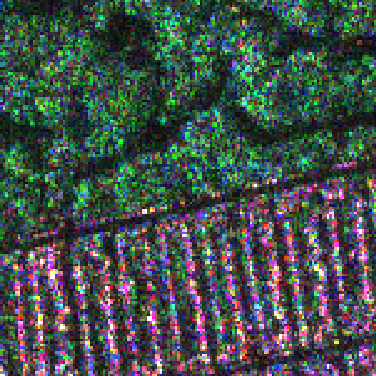}&
     \includegraphics[width=0.15\columnwidth]{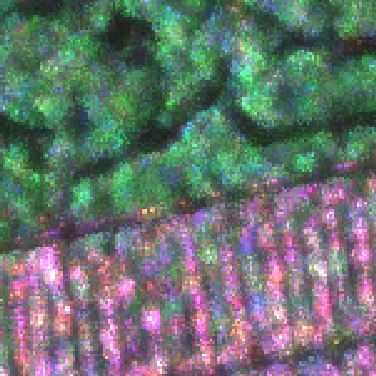}&
     \includegraphics[width=0.15\columnwidth]{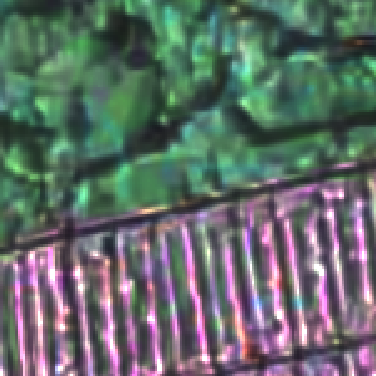}&
     \includegraphics[width=0.15\columnwidth]{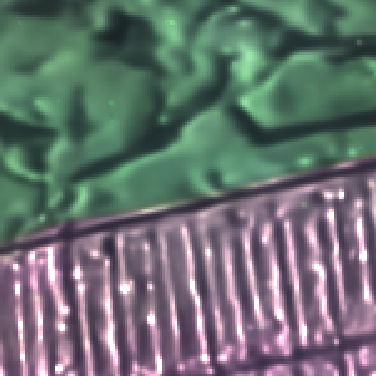}&
     \includegraphics[width=0.15\columnwidth]{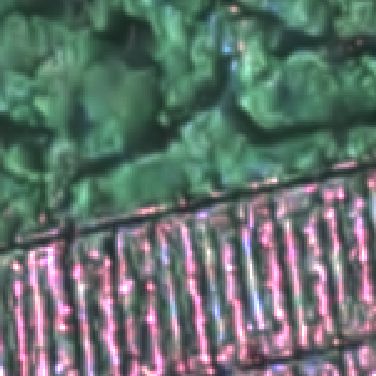}\\

\end{tabular}
    
    \caption{Close-up of the full-polarization results. From left to right the noisy image and filtering results. From top to bottom RADARSAT-2, BIOMASS and ALOS testing cases}
    \label{fig:full_comparison_zoom}
\end{figure}

\begin{table}[h]
\caption{Quantitative full-polarization despeckling results. The best value for each metric is highlighted in bold.}
\centering
\renewcommand{\arraystretch}{1.5}
\label{tab:full_pol_results}
\resizebox{\columnwidth}{!}{%
\begin{tabular}{llllll}
\hline
                                                  & Method   & KL-Tr $\downarrow$ & OVL-Tr $\uparrow$ & KL-$\lambda$max $\downarrow$ & OVL-$\lambda$max $\uparrow$ \\ \hhline{=|=====}
\multicolumn{1}{l|}{\multirow{4}{*}{\rotatebox[origin=c]{90}{RADARSAT-2}}}    & MEET-SAR & 1.3545             & 0.7364            & 11.3901                      & 0.6079                      \\
\multicolumn{1}{l|}{}                             & MuLog    & \textbf{0.0293}    & {\ul 0.9282}      & {\ul 0.0836}                 & {\ul 0.8761}                \\
\multicolumn{1}{l|}{}                             & DRUNet   & 0.2073             & 0.7333            & 0.1152                       & 0.7357                      \\
\multicolumn{1}{l|}{}                             & PANDA    & {\ul 0.0297}       & \textbf{0.9539}   & \textbf{0.0615}              & \textbf{0.8824}             \\ \hhline{=|=====}
\multicolumn{1}{l|}{\multirow{4}{*}{\rotatebox[origin=c]{90}{BIOMASS}}}     & MEET-SAR & 0.4693             & 0.7332            & 0.9943                       & 0.5769                      \\
\multicolumn{1}{l|}{}                             & MuLog    & 0.0766             & 0.8883            &  0.1131              & \underline{0.8398}                      \\
\multicolumn{1}{l|}{}                             & DRUNet   & {\ul 0.0250}       & {\ul 0.9171}      & \textbf{0.0089}              & \textbf{0.9526}             \\
\multicolumn{1}{l|}{}                             & PANDA    & \textbf{0.0146}    & \textbf{0.9428}   & \underline{0.1044}                       & 0.8232              \\ \hhline{=|=====}
\multicolumn{1}{l|}{\multirow{4}{*}{\rotatebox[origin=c]{90}{ALOS}}} & MEET-SAR & 0.0999             & 0.8865            & 1.5256                       & 0.7473                      \\
\multicolumn{1}{l|}{}                             & MuLog    & 0.0203             & 0.9374            & 0.0927                       & 0.8662                      \\
\multicolumn{1}{l|}{}                             & DRUNet   & {\ul 0.0079}       & {\ul 0.9595}      & \textbf{0.0377}              & \textbf{0.8873}             \\
\multicolumn{1}{l|}{}                             & PANDA    & \textbf{0.0043}    & \textbf{0.9832}   & {\ul 0.0501}                 & {\ul 0.8765}                \\ \hhline{=|=====}
\end{tabular}%
}
\end{table}
\begin{figure}[h]
    \centering
    \begin{tabular}{c}
    RADARSAT-2\\
        \includegraphics[width=0.6\columnwidth]{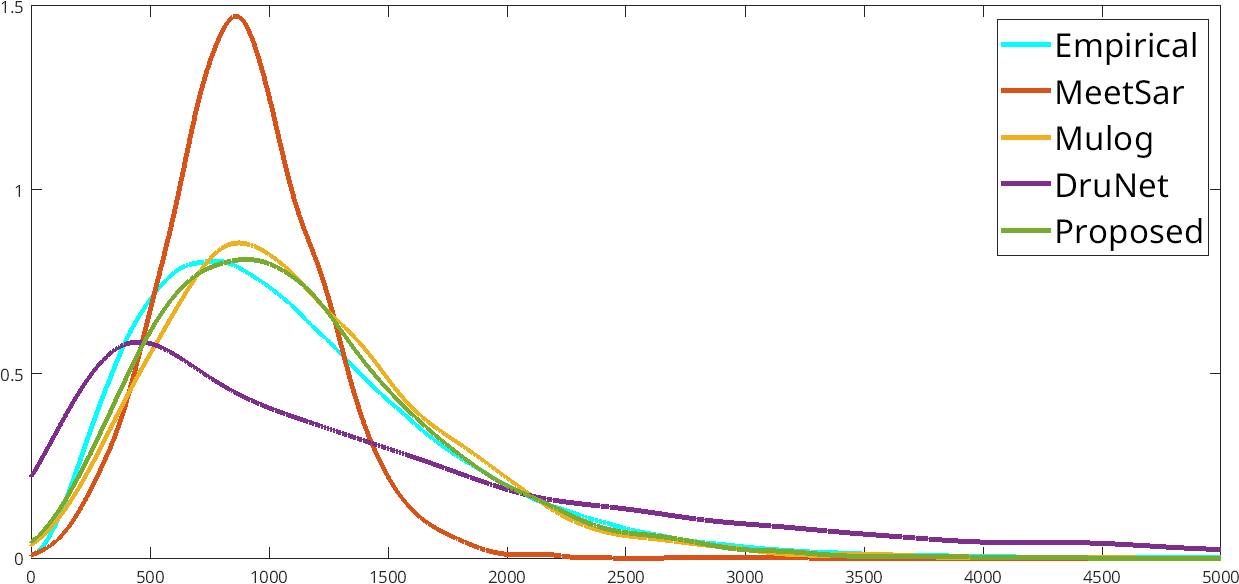} \\
        \includegraphics[width=0.6\columnwidth]{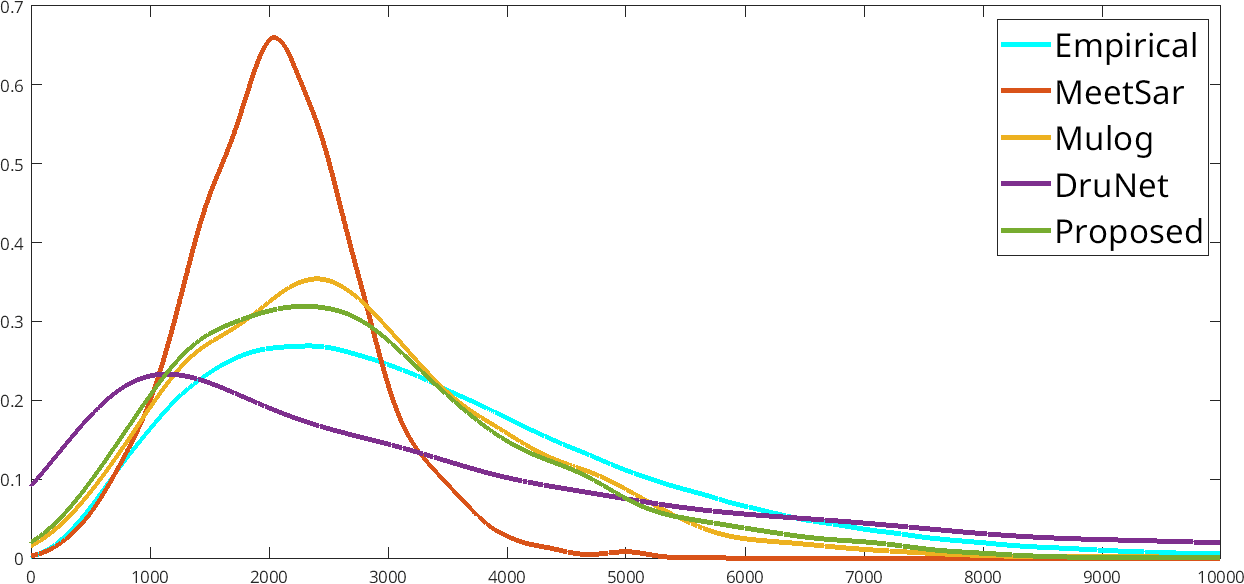} \\
        BIOMASS \\\includegraphics[width=0.6\columnwidth]{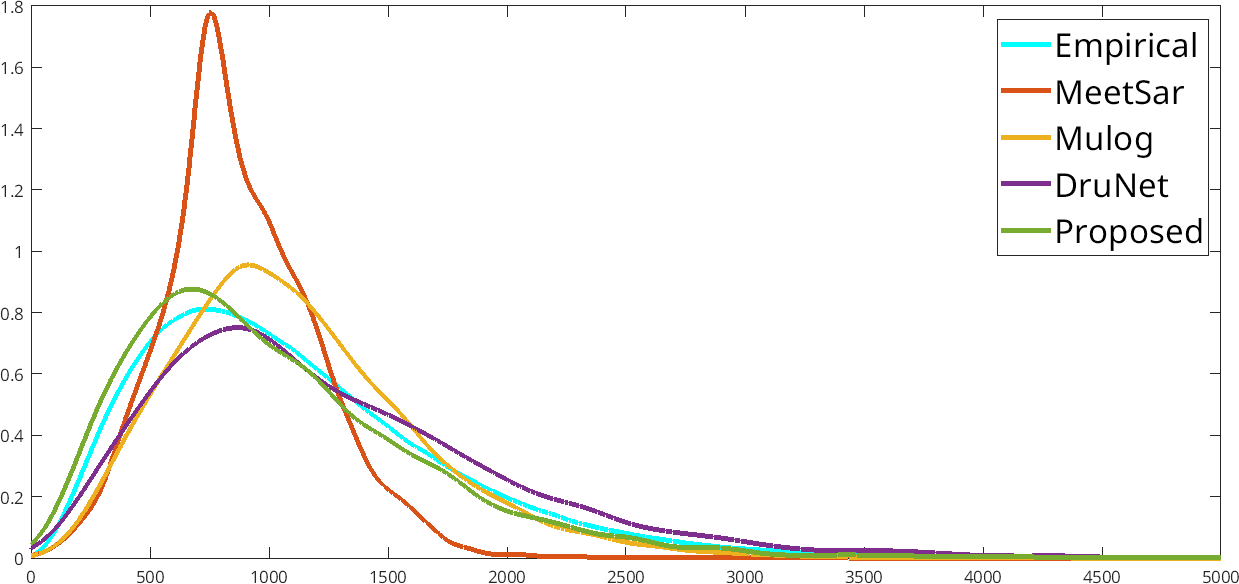}\\
        \includegraphics[width=0.6\columnwidth]{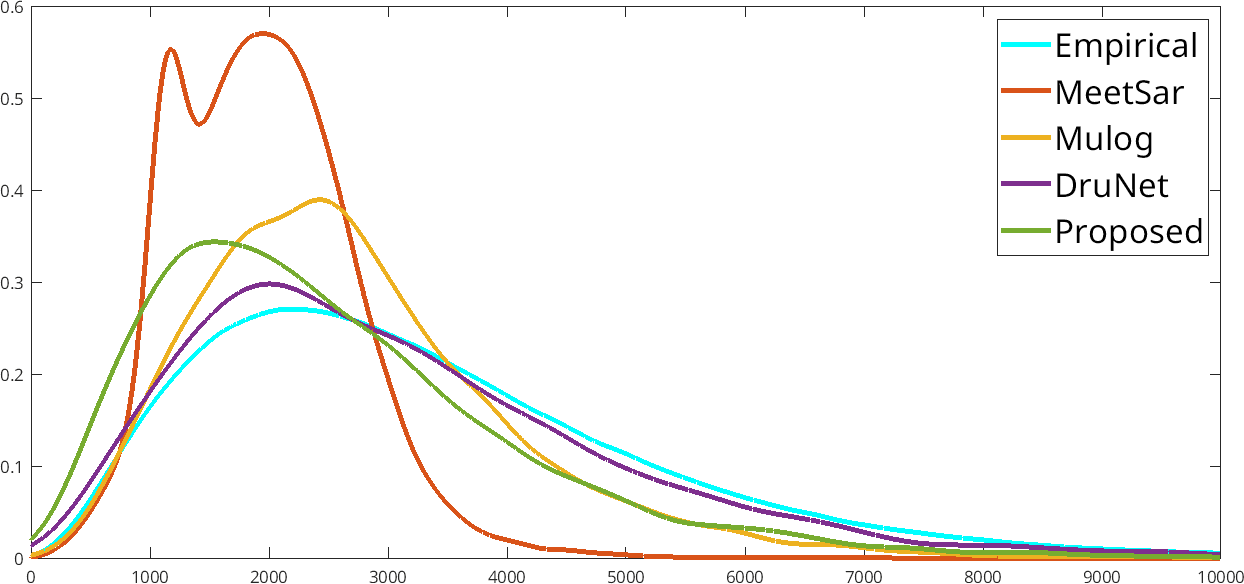} \\
        ALOS PALSAR\\
        \includegraphics[width=0.6\columnwidth]{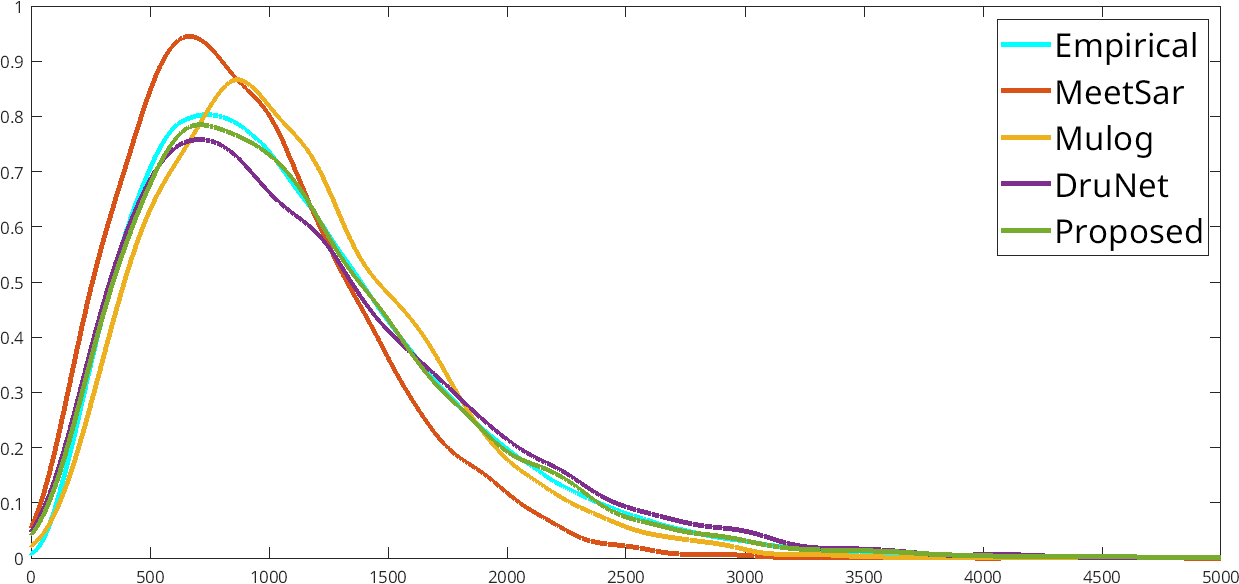}\\
\includegraphics[width=0.6\columnwidth]{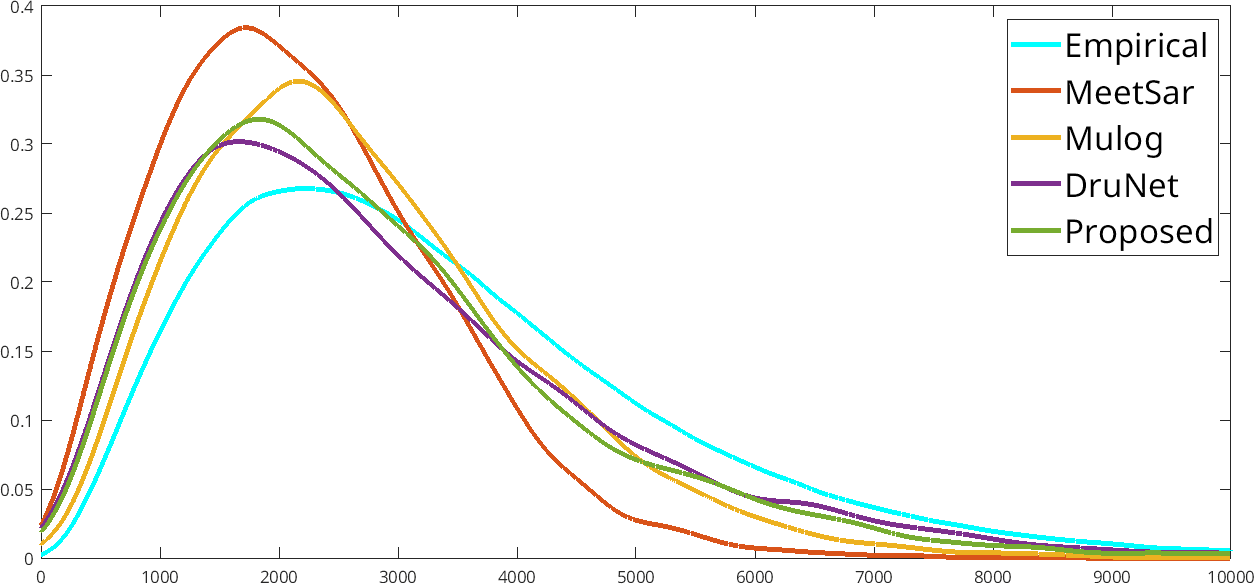} \\
        \end{tabular}
    \caption{Residual speckle PDF analysis for the full-polarimetric case. For each sensor, normalized trace in the top and largest eigenvalue  in the bottom.}
    \label{fig:full_pdf}
\end{figure}

More specifically, this is  more clear observing the close-ups shown in Figure \ref{fig:full_comparison_zoom}: \prop{} is able to better preserve tiny details like the roads (in RADARASAT-2) or fields separation (in BIOMASS) and vegetation structure ( in ALOS) that are instead almost lost in DRUNet and MEET-SAR. Instead MuLog has a better polarimetric preservation at the cost of a presence of some artifacts. 

These comments find confirm from quantitative analysis. The statistical consistency described in Section \ref{subsec:evaluation_metrics} has been performed over homogeneous patches of dimension $100 \times 100$ pixels: visual comparison of the estimated distribution with the theoretical one is reported in Figure \ref{fig:full_pdf}, while quantitative analysis in Table \ref{tab:full_pol_results}. \prop{} achieves the best performance on the preservation of the statistics of the trace of the residual covariance matrix except for the KL on RADARSAT-2 where achieves the second one. This indicates a good speckle suppression and spatial preservation. Instead, the preservation of statistics of the eigenvalue from the cross-polarized elements of the residual covariance matrix achieves similar performance of MuLog, proving the effectiveness on polarimetric preservation. It is worthy to note that metrics highlight BIOMASS to be the more challenging area for the preservation of polarimetric mechanism for \prop{}: this is mostly due to the the presence of water and sea that are not enough represented in the original training dataset.

\subsection{Dual-Polarization Results}

\begin{figure*}[t]
    \centering
    \begingroup
    \newsavebox{\dualfirstpanel}
    \newcommand{\duallabelledpanel}[2]{%
        \sbox{\dualfirstpanel}{%
            \includegraphics[width=0.15\textwidth]{#2}}%
        \makebox[0pt][r]{%
            \smash{\raisebox{\dimexpr\ht\dualfirstpanel/2\relax}{%
                \raisebox{-0.5\height}{%
                    \rotatebox{90}{\small\textbf{#1}}}}}%
            \hspace{4pt}}%
        \usebox{\dualfirstpanel}}

    \begin{tabular}{ccccc}
        SAR & MEET-SAR & MuLog & CNN-PolSAR & \prop{} \\

        \duallabelledpanel{RADARSAT-2}{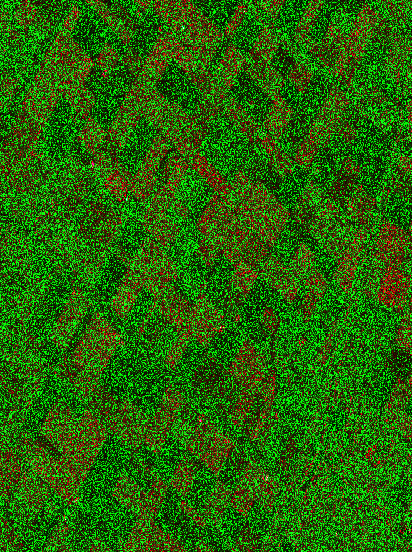}&
        \includegraphics[width=0.15\textwidth]{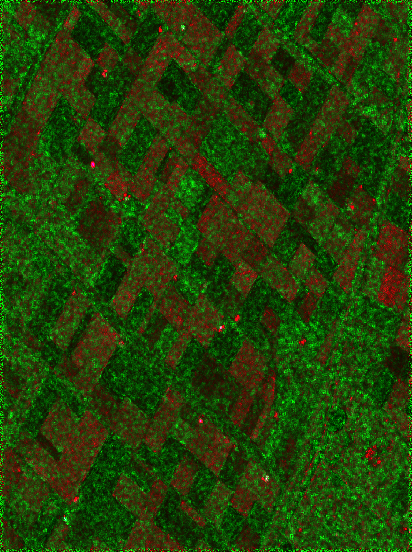}&
        \includegraphics[width=0.15\textwidth]{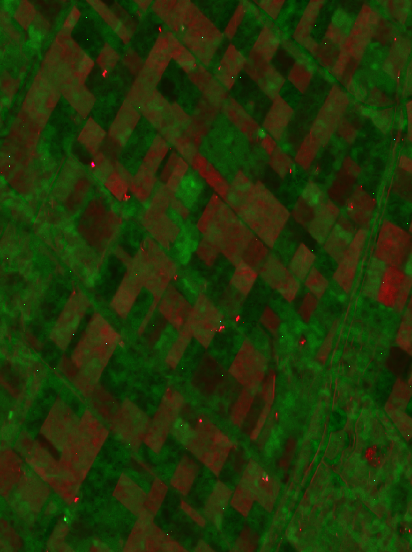}&
        \includegraphics[width=0.15\textwidth]{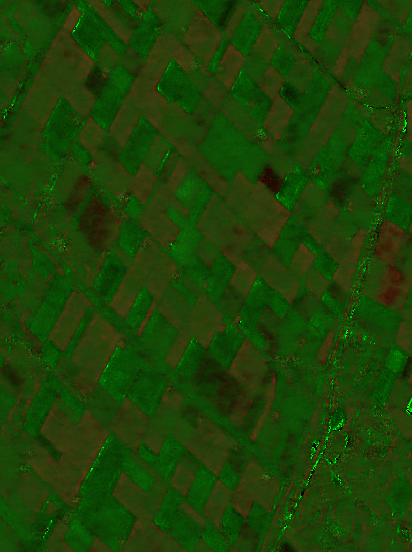}&
        \includegraphics[width=0.15\textwidth]{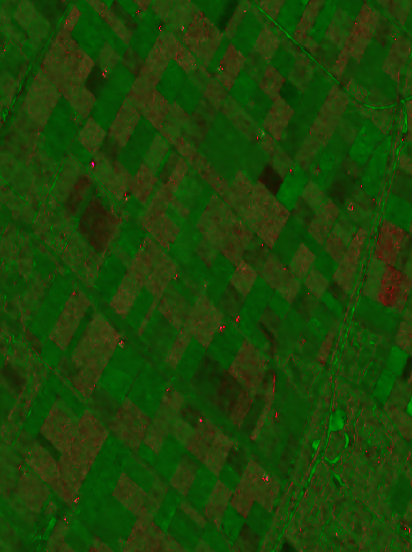}\\

        \duallabelledpanel{BIOMASS}{Images/DualPol/BIOMASS_dualpol_noisy.png}&
        \includegraphics[width=0.15\textwidth]{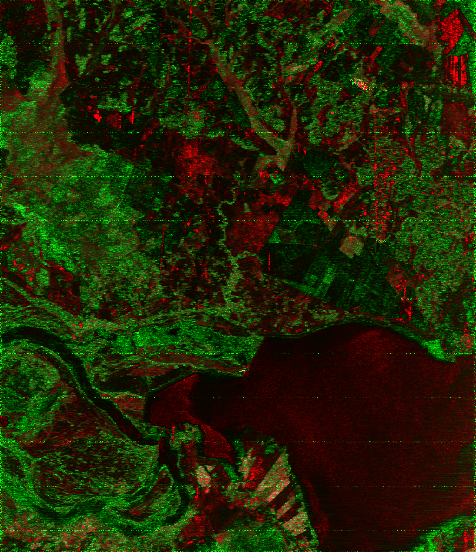}&
        \includegraphics[width=0.15\textwidth]{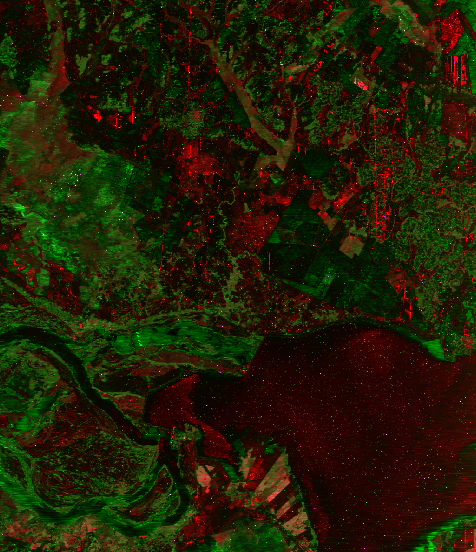}&
        \includegraphics[width=0.15\textwidth]{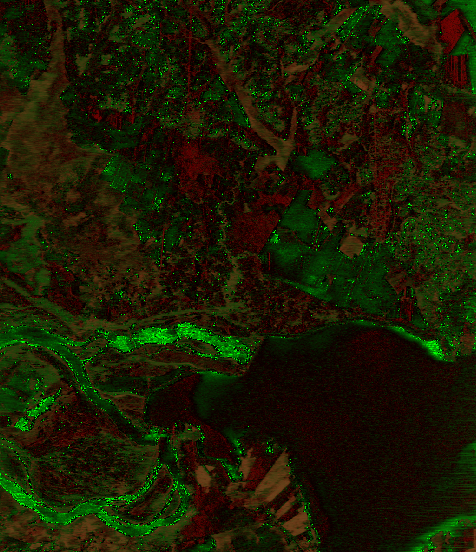}&
        \includegraphics[width=0.15\textwidth]{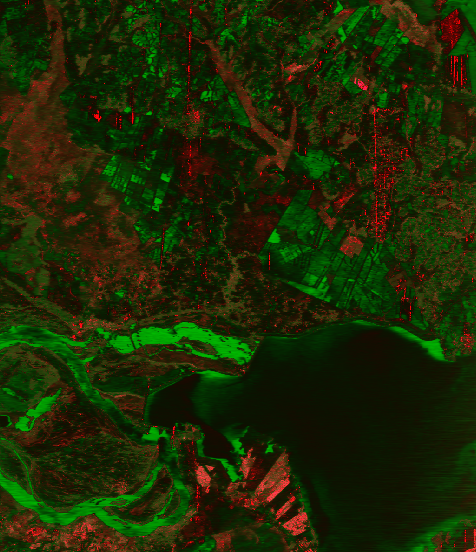}\\

        \duallabelledpanel{ALOS PALSAR}{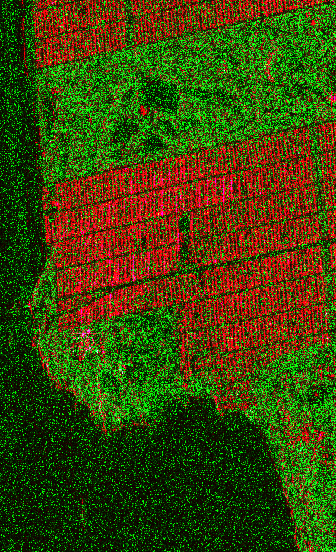}&
        \includegraphics[width=0.15\textwidth]{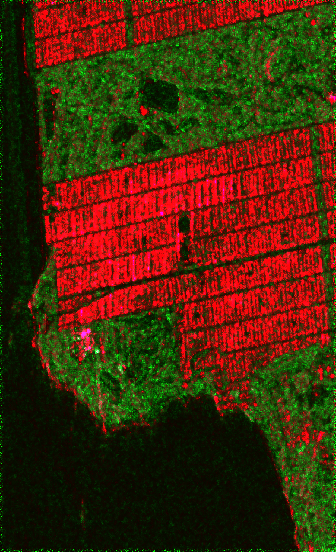}&
        \includegraphics[width=0.15\textwidth]{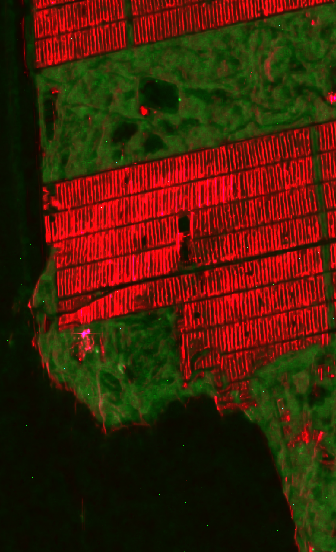}&
        \includegraphics[width=0.15\textwidth]{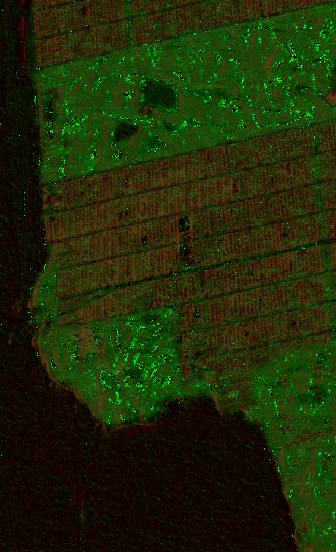}&
        \includegraphics[width=0.15\textwidth]{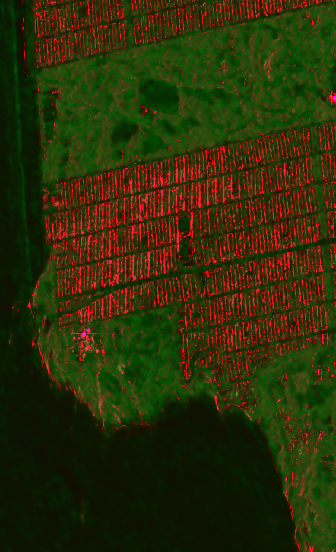}\\
    \end{tabular}
    \endgroup

    \caption{Dual-polarization results. From left to right the noisy image and filtering results. From top to bottom RADARSAT-2, BIOMASS and ALOS testing cases. The results are displayed in RGB composition with R=HH, G=HV/HH, B=HV}
    \label{fig:dual_comparison}
\end{figure*}

As for the full-polarization case, two model based solutions and a deep learning one have been selected as comparison methods: MEET-SAR, MuLog, and CNN-PolSAR. The qualitative assessment over the three test cases is shown in Figure \ref{fig:dual_comparison}:  from left to right the noisy image, MEET-SAR, MuLog, CNN-PolSAR and \prop{}; from top to bottom results on RADARSAT-2, BIOMASS and ALOS-PALSAR. The images are represented in RGB composition with R=HH channel, G= HV/HH ratio and B=HV channel: in this way the red channel should highlight the surface scattering mechanism (i.e. urban, buildings) while the green the volume one (i.e. vegetation, forest, agricultural fields).

\begin{figure}[h]
     \centering
     \begin{tabular}{ccccc}
     SAR & MEET-SAR & MuLog & CNN & \prop{}\\
    \includegraphics[width=0.15\columnwidth]{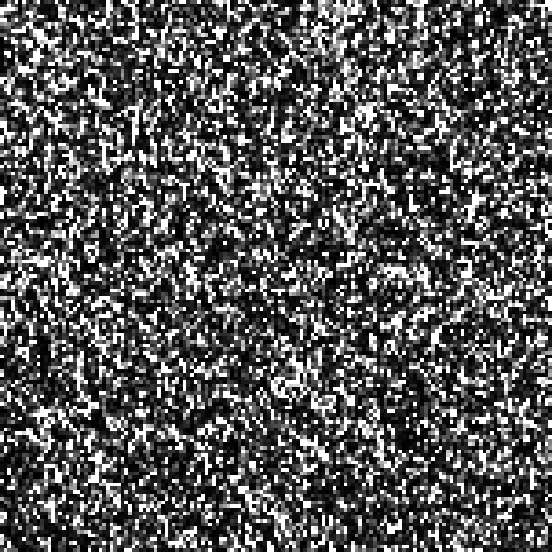}&
     \includegraphics[width=0.15\columnwidth]{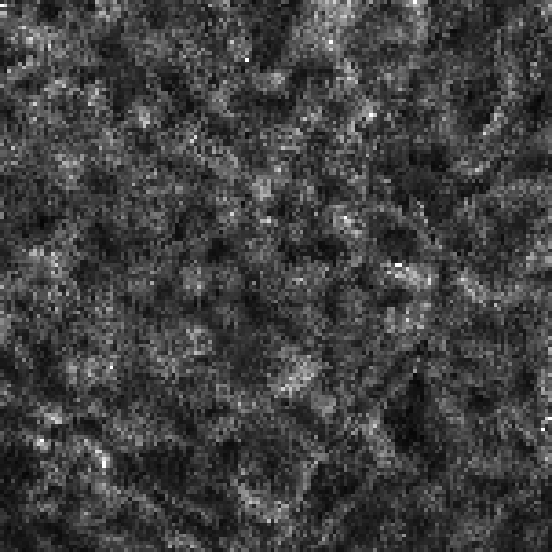}&
     \includegraphics[width=0.15\columnwidth]{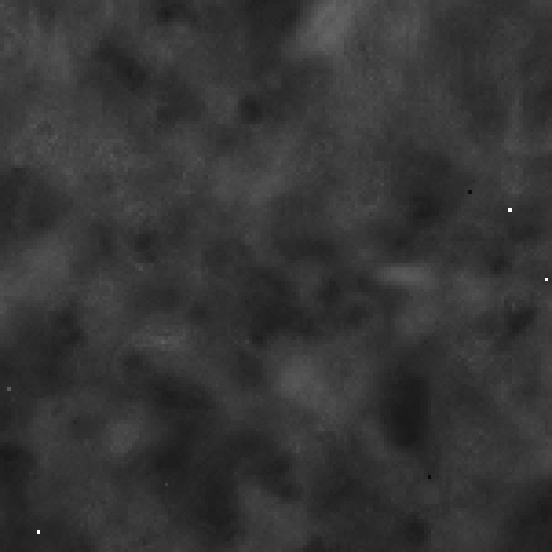}&
     \includegraphics[width=0.15\columnwidth]{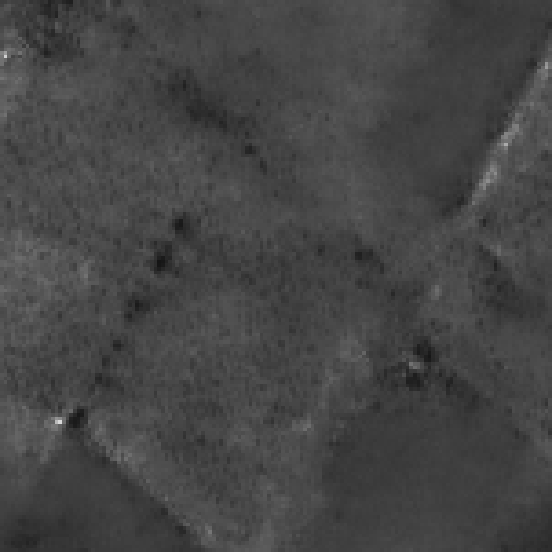}&
     \includegraphics[width=0.15\columnwidth]{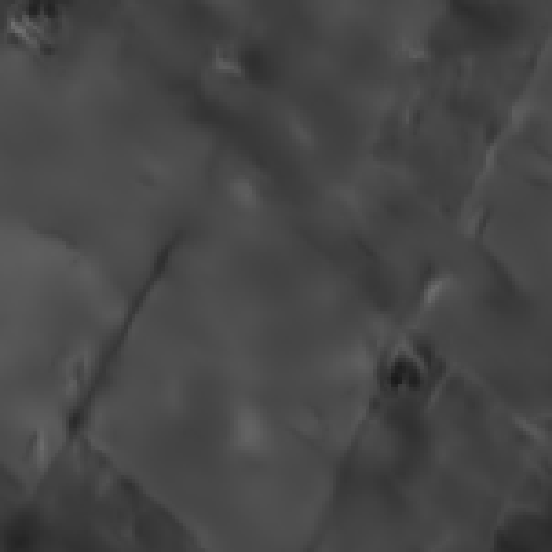}\\
    
     \includegraphics[width=0.15\columnwidth]{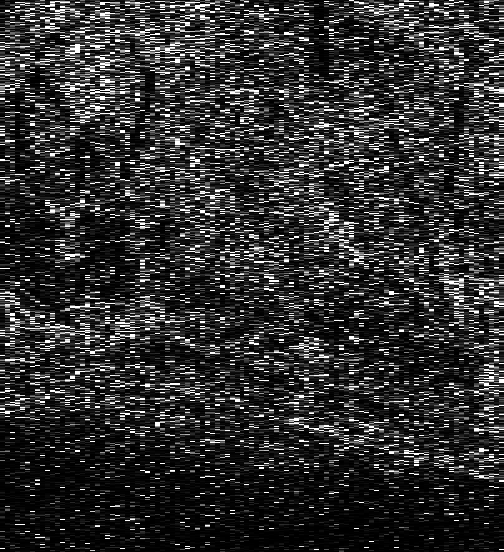}&
     \includegraphics[width=0.15\columnwidth]{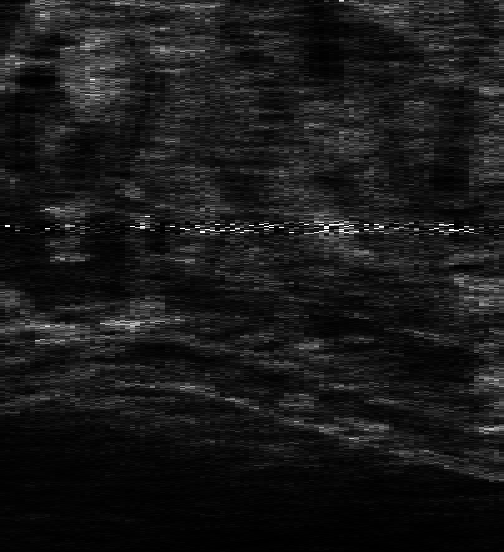}&
     \includegraphics[width=0.15\columnwidth]{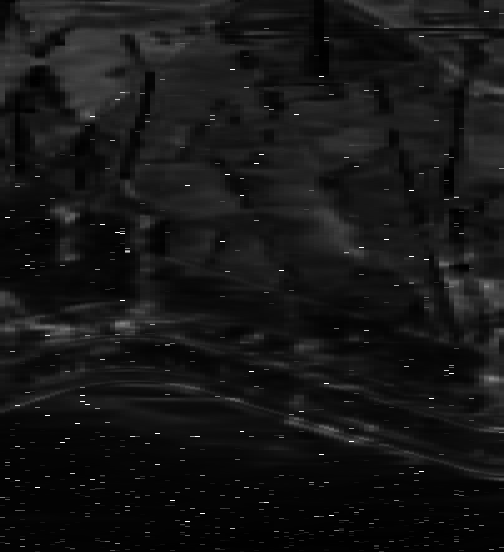}&
     \includegraphics[width=0.15\columnwidth]{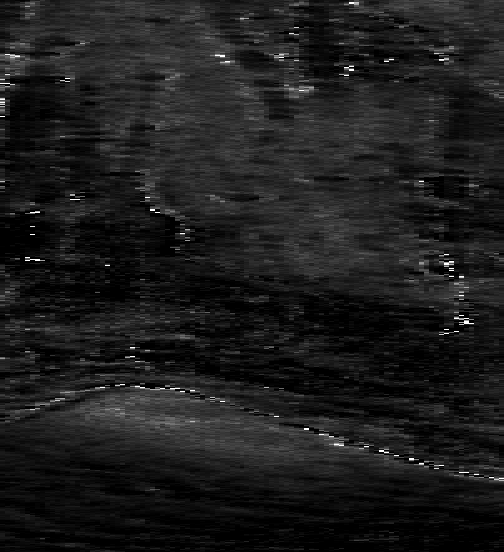}&
     \includegraphics[width=0.15\columnwidth]{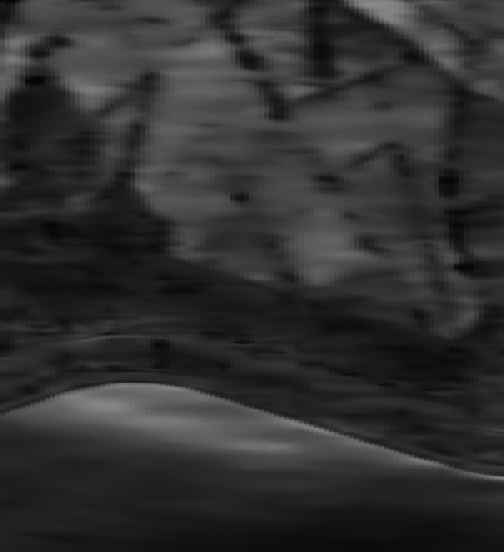}\\

     \includegraphics[width=0.15\columnwidth]{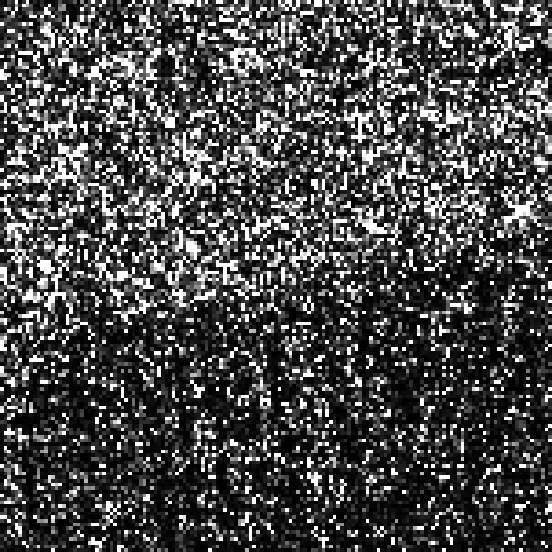}&
     \includegraphics[width=0.15\columnwidth]{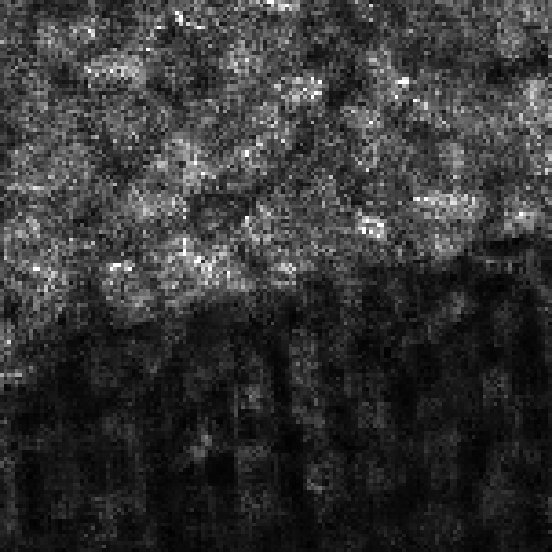}&
     \includegraphics[width=0.15\columnwidth]{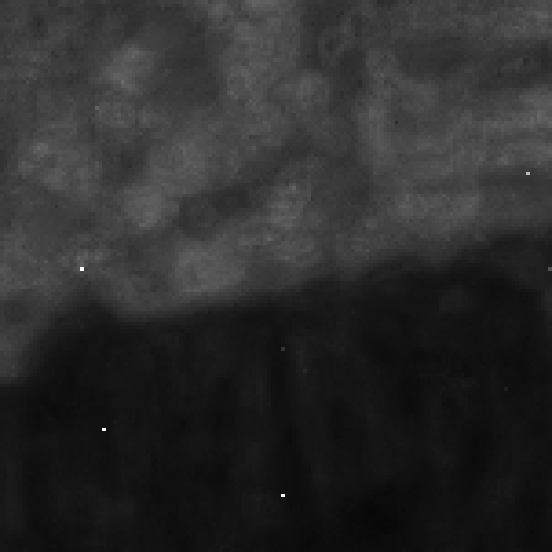}&
     \includegraphics[width=0.15\columnwidth]{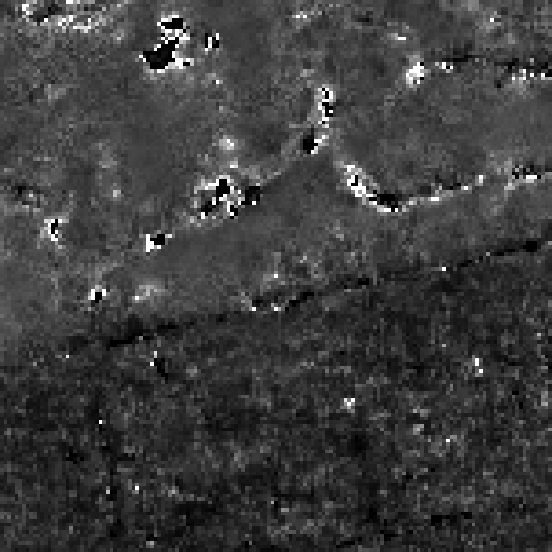}&
     \includegraphics[width=0.15\columnwidth]{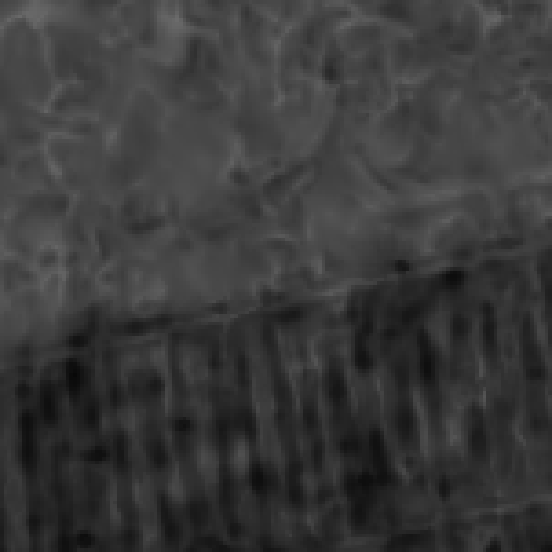}\\
     \end{tabular}
    
    \caption{Close-up of the ratio HV/HH of dual-polarization results. From left to right the noisy image and filtering results. From top to bottom RADARSAT-2, BIOMASS and ALOS testing cases.}
    \label{fig:dual_comparison_zoom}
\end{figure}

\begin{figure}[h]
    \centering
    \begin{tabular}{c}
    RADARSAT-2\\
        \includegraphics[width=0.6\columnwidth]{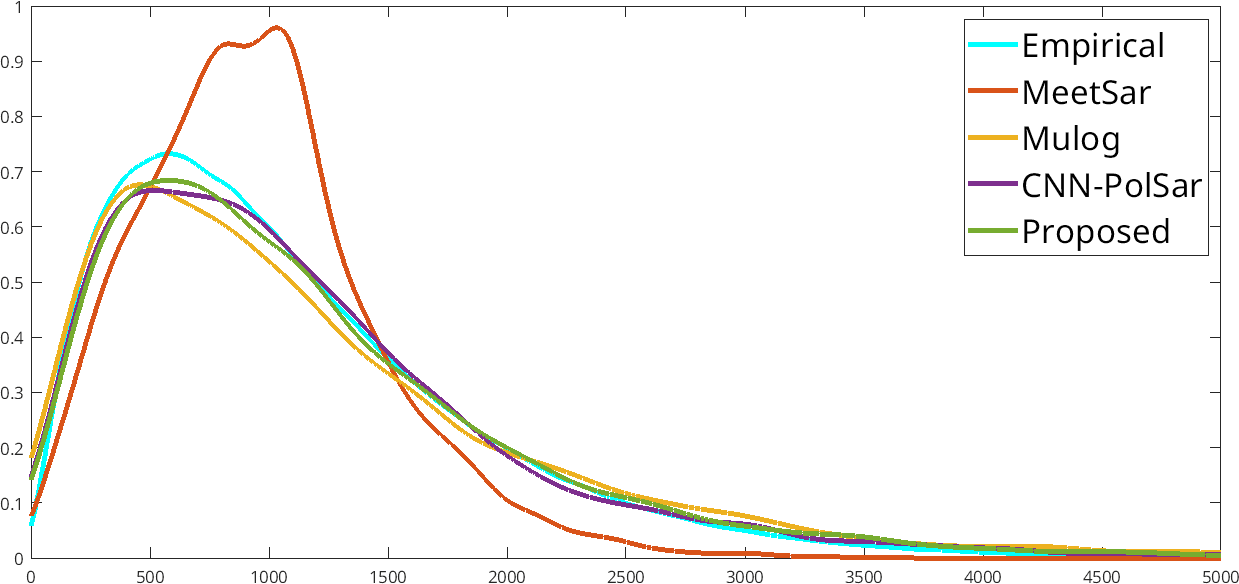} \\
        \includegraphics[width=0.6\columnwidth]{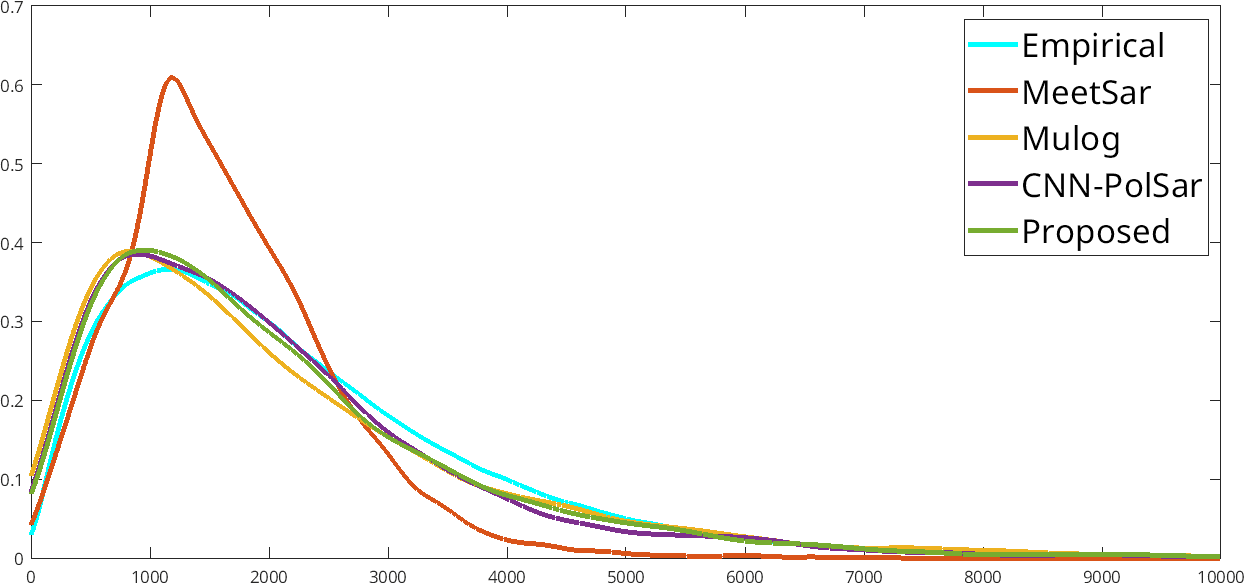} \\
        BIOMASS \\
        \includegraphics[width=0.6\columnwidth]{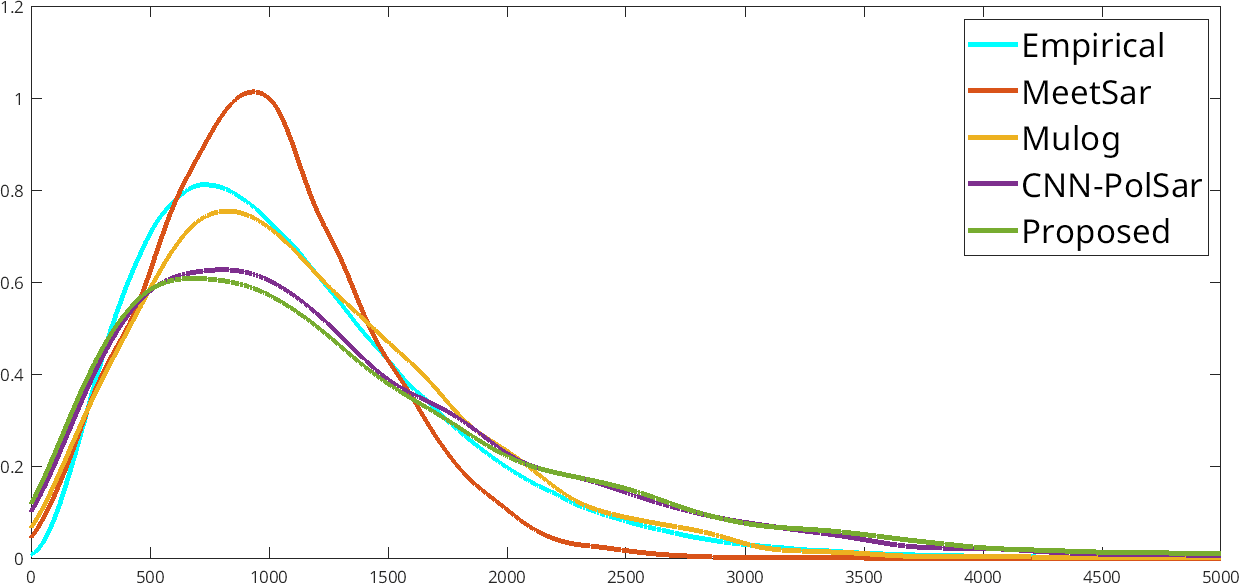}\\
        \includegraphics[width=0.6\columnwidth]{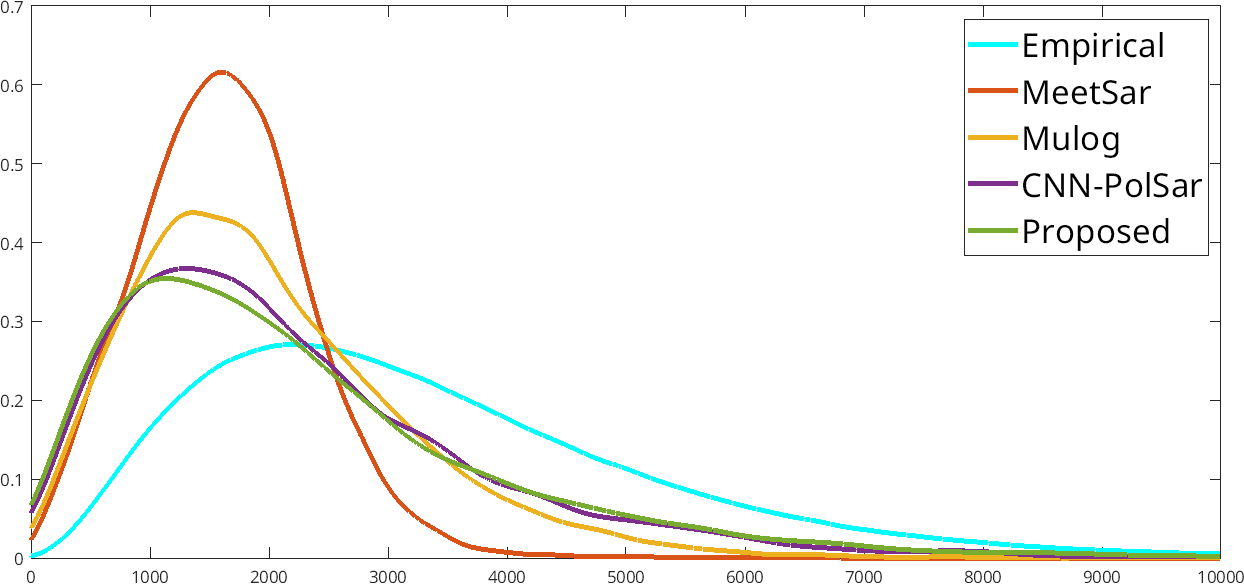} \\
        ALOS PALSAR\\
        \includegraphics[width=0.6\columnwidth]{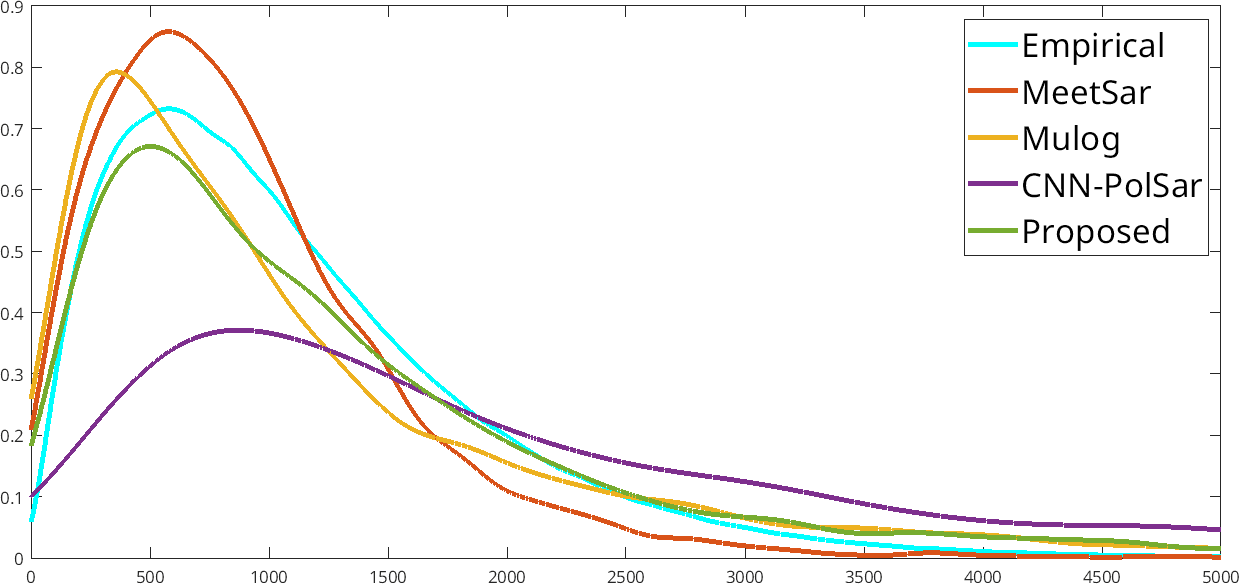}\\
\includegraphics[width=0.6\columnwidth]{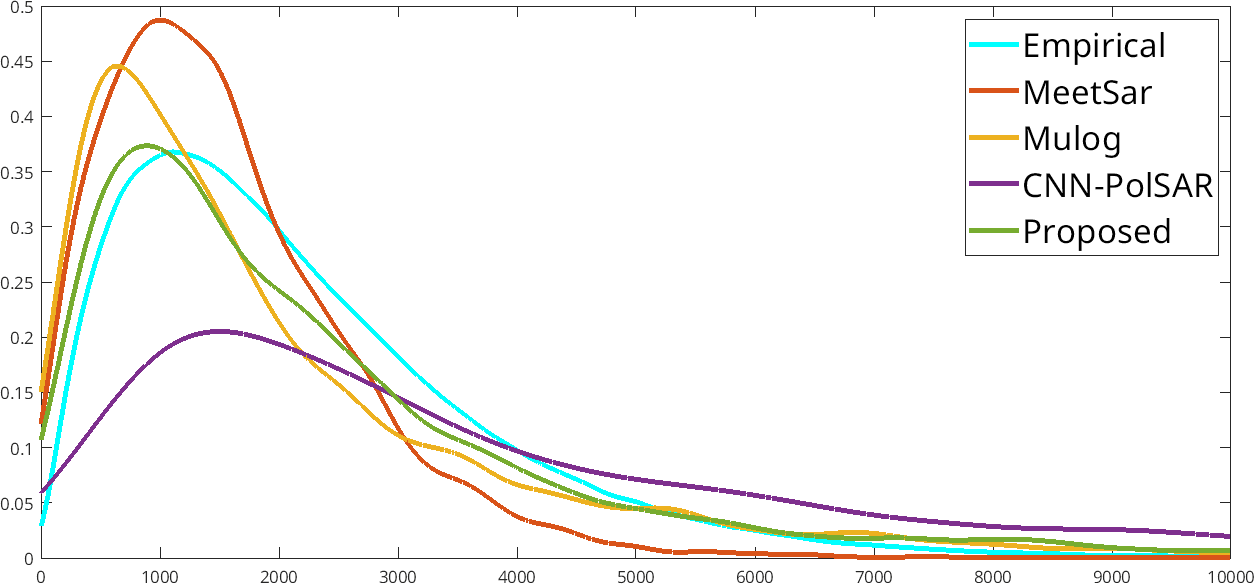} \\
        \end{tabular}
    \caption{Residual speckle PDF analysis for the dual-polarimetric case. For each sensor, normalized trace in the top and largest eigenvalue  in the bottom.}
    \label{fig:dual_pdf}
\end{figure}

From the Figure \ref{fig:dual_comparison} comments similar to the full-polarization case can be reported for MEET-SAR, MuLog and \prop{}: MEET-SAR shows a good polarimetric scattering preservation but a limited speckle reduction, MuLog shows best performance in urban area preservation and good balance between speckle reduction, spatial details and polarimetric scattering preservation with a bit of over-smoothing of the vegetated areas. Indeed, while MuLog strongly performs on ALOS with a good separation between vegetation and urban, it shows some polarimetric distortions in not surface areas (red zones appear in agricultural fields of RADARSAT-2 or many red spot on water/sea of BIOMASS). \prop{} shows less ability than MuLog to preserve spatial details in the urban area, but shows a robust speckle suppression, spatial and polarimetric preservation over the vegetated area. This behaviour is quite general in the three testing cases. CNN-PolSAR instead is not able to well generalize on the three different sensors: it shows good performance on RADARSAT-2 while the speckle suppression and polarimetric scattering mechanism are less preserved in the other two cases. In order to have a better evidence of these properties, we reported a close-up of the only ratio HV/HH in Figure \ref{fig:dual_comparison_zoom}: it is evident that \prop{} is able to better preserve the details in vegetation areas of RADARSAT-2 and BIOMASS rather than the other methods. At the same time, it is also evident the ability of MuLog in ensuring a better separation of volume and scattering mechanisms in the ALOS dataset. These comments are confirmed from visual comparison of the estimated distribution with
the theoretical one reported in Figure \ref{fig:dual_pdf} and from the quantitative analysis in Table \ref{tab:dual_pol} where the statistical consistency described in Section \ref{subsec:evaluation_metrics} is reported. It worthy to note that, in this case with less polarization information with respect the full-polarized one, \prop{} seems to well generalized on areas not well represented on the training dataset (i.e. the water and sea of BIOMASS area).

\begin{table}[h]
\caption{Quantitative dual-polarization despeckling results. The best value for each metric is highlighted in bold.}
\label{tab:dual_pol}
\centering
\renewcommand{\arraystretch}{1.5}
\resizebox{\columnwidth}{!}{%
\begin{tabular}{llllll}
\hline
                                                  & Method   & KL-Tr $\downarrow$ & OVL-Tr $\uparrow$ & KL-$\lambda$max $\downarrow$ & OVL-$\lambda$max $\uparrow$ \\ \hhline{=|=====}
\multicolumn{1}{l|}{\multirow{4}{*}{\rotatebox[origin=c]{90}{RADARSAT-2}}}    & 
MEET-SAR & 0.9571 & 0.8344 & 0.8611 & 0.8044 \\
\multicolumn{1}{l|}{} & MuLog & 0.0160 & 0.9348 & 0.0145 & 0.9254 \\
\multicolumn{1}{l|}{} & CNN-PolSAR & \underline{0.0090} & \underline{0.9614} & \underline{0.0115} & \underline{0.9441} \\
\multicolumn{1}{l|}{} & PANDA & \textbf{0.0079} & \textbf{0.9621} &
\textbf{0.0079} & \textbf{0.9454} \\
\hhline{=|=====}
\multicolumn{1}{l|}{\multirow{4}{*}{\rotatebox[origin=c]{90}{BIOMASS}}}     & 
MEET-SAR & 0.3855 & \underline{0.8810} & 6.6587 & 0.5026 \\
\multicolumn{1}{l|}{} & MuLog & \textbf{0.0117} & \textbf{0.9481} &
0.5034 & 0.6745 \\
\multicolumn{1}{l|}{} & CNN-PolSAR & \underline{0.0564} & 0.8750 & \underline{0.2145} & \underline{0.7319} \\
\multicolumn{1}{l|}{} & PANDA & 0.0751 & 0.8563 &
\textbf{0.1942} & \textbf{0.7388} \\
\hhline{=|=====}
\multicolumn{1}{l|}{\multirow{4}{*}{\rotatebox[origin=c]{90}{ALOS}}} &
MEET-SAR & 0.0636 & \underline{0.8849} & 0.1572 & 0.8161 \\
\multicolumn{1}{l|}{} & MuLog & \underline{0.0609} & 0.8572 & \underline{0.0655} & \underline{0.8354} \\
\multicolumn{1}{l|}{} & CNN-PolSAR & 0.2904 & 0.6457 & 0.1392 & 0.6802 \\
\multicolumn{1}{l|}{} & PANDA & \textbf{0.0459} & \textbf{0.9012} &
\textbf{0.0266} & \textbf{0.8912} \\
\hhline{=|=====}
\end{tabular}%
}
\end{table}

% \begin{figure*}[t]
%     \centering
%     \begin{tabular}{ccc}
%     RADARSAT-2 & BIOMASS & ALOS PALSAR \\
%         \includegraphics[width=0.3\textwidth]{Images/DualPol/radarsat_dualpol_pdf_trace.png} & \includegraphics[width=0.3\textwidth]{Images/DualPol/BIOMASS_dualpol_pdf_trace.png}& \includegraphics[width=0.3\textwidth]{Images/DualPol/alos_dualpol_pdf_trace.png}\\
%         \includegraphics[width=0.3\textwidth]{Images/DualPol/radarsat_dualpol_pdf_eig.png} & \includegraphics[width=0.3\textwidth]{Images/DualPol/alos_dualpol_pdf_eig.png} & \includegraphics[width=0.3\textwidth]{Images/DualPol/alos_dualpol_pdf_eig.png}\\
%         \end{tabular}
%     \caption{Residual speckle PDF analysis for the dual-polarimetric case. Normalized trace in the top and largest eigenvalue  in the bottom.}
%     \label{fig:dual_pdf}
% \end{figure*}

\subsection{Single-Polarization Results}
\label{subsec:single_pol_results}

\begin{figure*}[h]
    \centering
    \setlength{\tabcolsep}{3pt}
    \renewcommand{\arraystretch}{1.1}

    \begin{tabular}{cccccc}
        & SAR & MEET-SAR & MuLog & CL-SAR & \prop{} \\

        \raisebox{0.5\height}{\rotatebox{90}{\textbf{RADARSAT-2}}} &
        \includegraphics[width=0.15\textwidth]{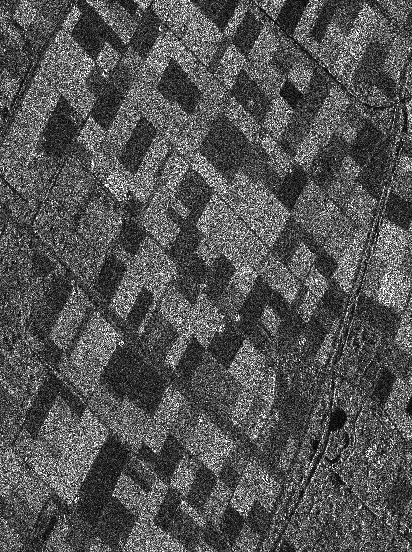} &
        \includegraphics[width=0.15\textwidth]{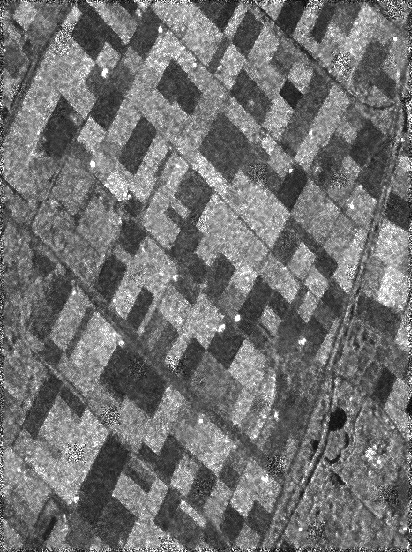} &
        \includegraphics[width=0.15\textwidth]{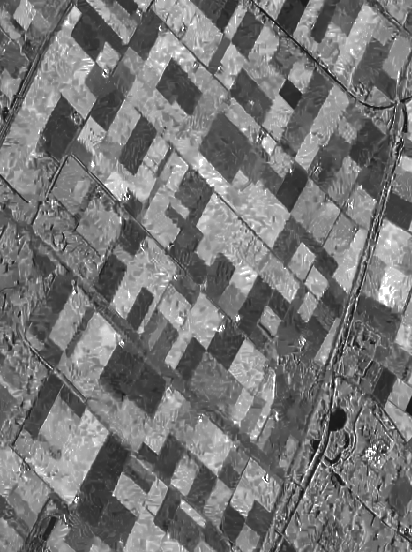} &
        \includegraphics[width=0.15\textwidth]{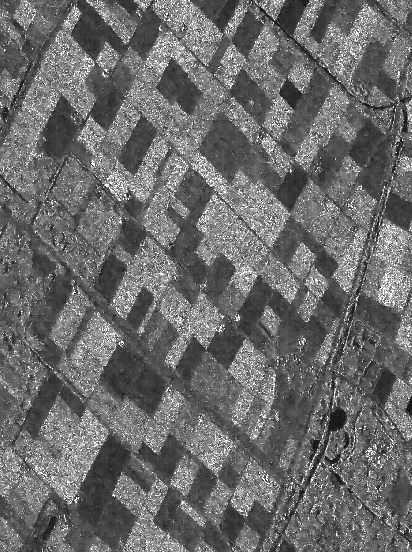} &
        \includegraphics[width=0.15\textwidth]{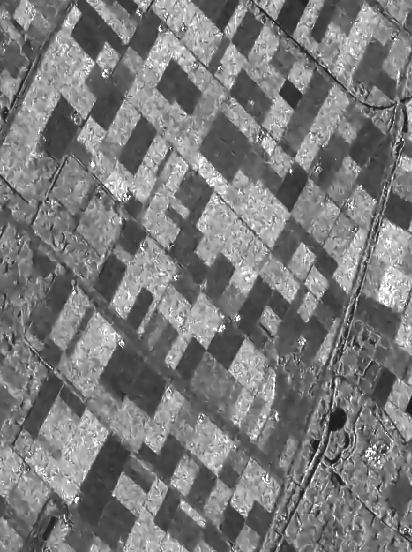} \\

        \raisebox{0.5\height}{\rotatebox{90}{\textbf{BIOMASS}}} &
        \includegraphics[width=0.15\textwidth]{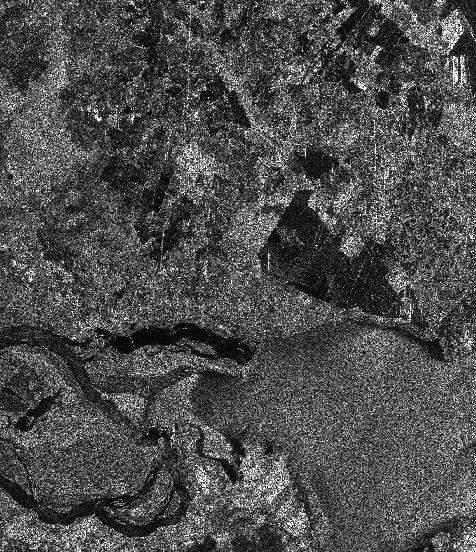} &
        \includegraphics[width=0.15\textwidth]{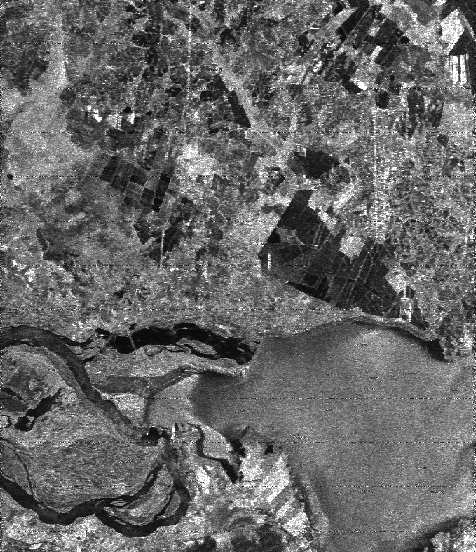} &
        \includegraphics[width=0.15\textwidth]{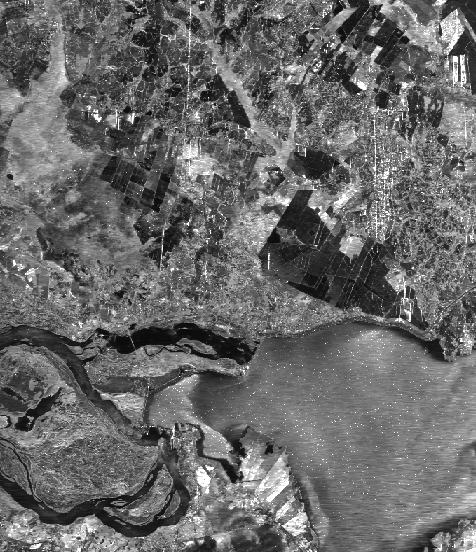} &
        \includegraphics[width=0.15\textwidth]{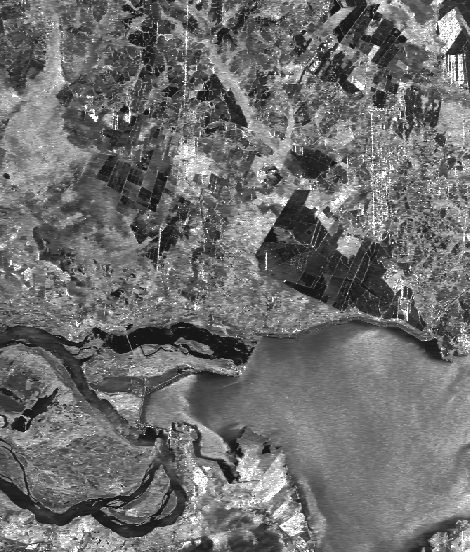} &
        \includegraphics[width=0.15\textwidth]{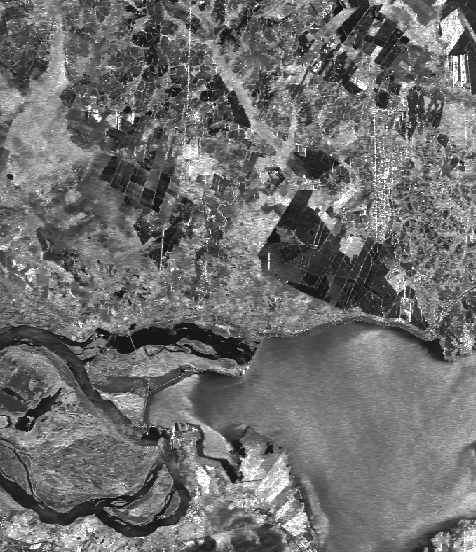} \\

        \raisebox{0.5\height}{\rotatebox{90}{\textbf{ALOS-PALSAR}}} &
        \includegraphics[width=0.15\textwidth]{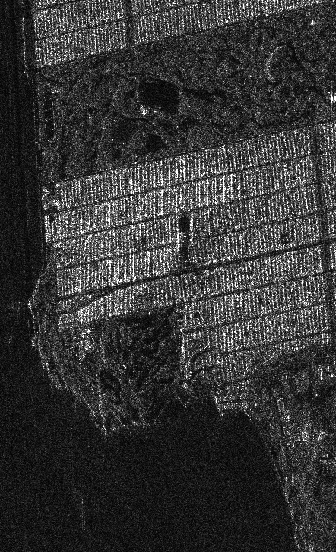} &
        \includegraphics[width=0.15\textwidth]{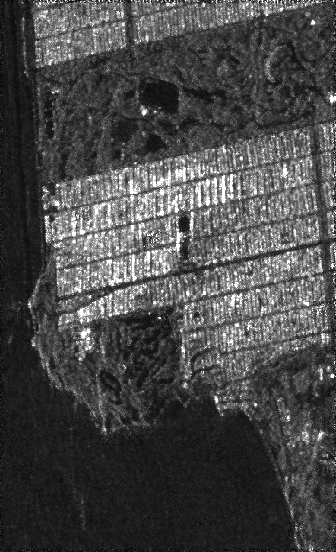} &
        \includegraphics[width=0.15\textwidth]{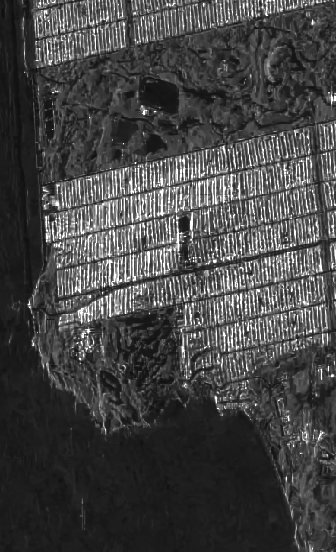} &
        \includegraphics[width=0.15\textwidth]{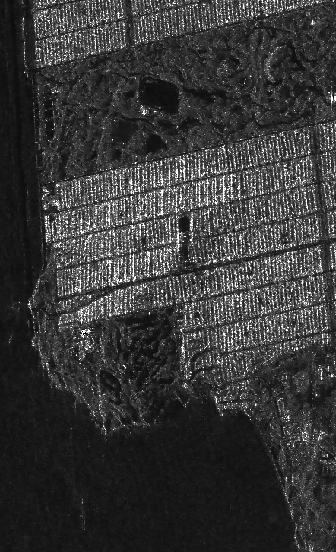} &
        \includegraphics[width=0.15\textwidth]{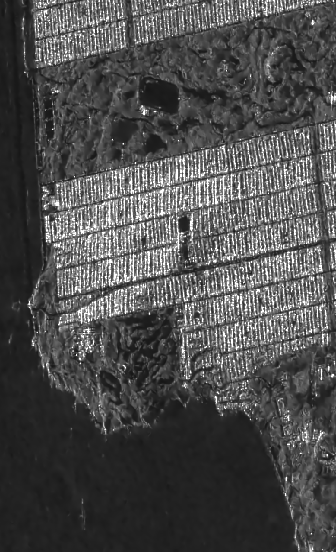} \\

    \end{tabular}

    \caption{Single-polarization results. From left to right the noisy image and filtering results. From top to bottom RADARSAT-2, BIOMASS and ALOS testing cases.}
    \label{fig:single_comparison}
\end{figure*}

\begin{figure}[h]
     \centering
     \begin{tabular}{ccccc}
     SAR & MEET-SAR & MuLog & CL-SAR & \prop{}\\
     
     \includegraphics[width=0.15\columnwidth]{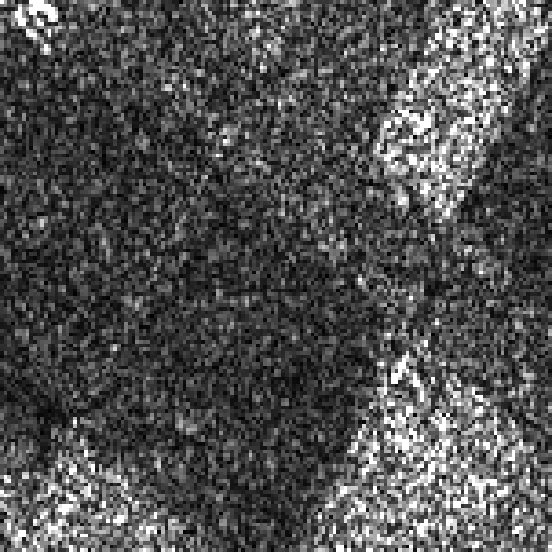}&
     \includegraphics[width=0.15\columnwidth]{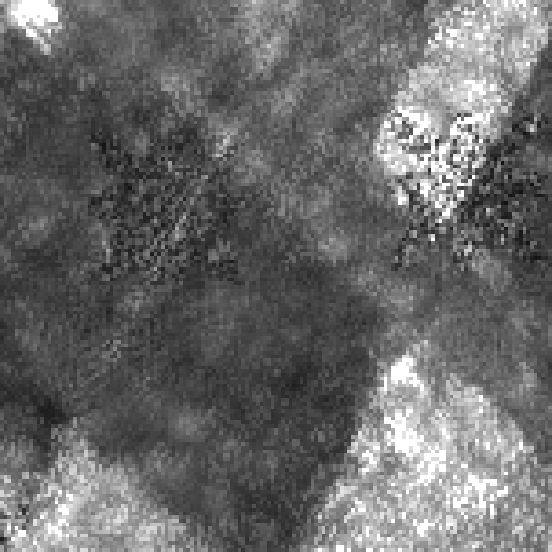}&
     \includegraphics[width=0.15\columnwidth]{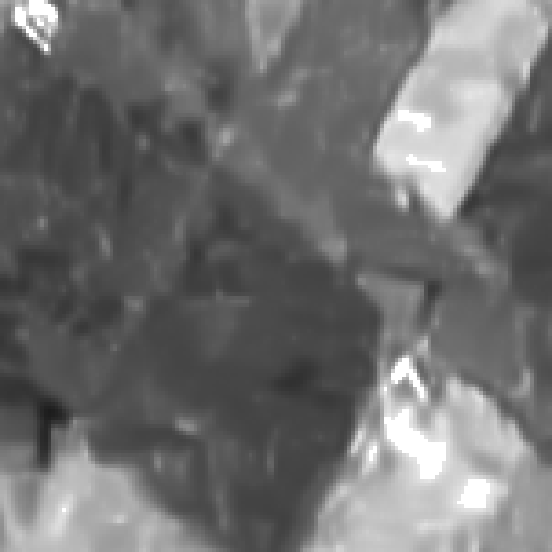}&
     \includegraphics[width=0.15\columnwidth]{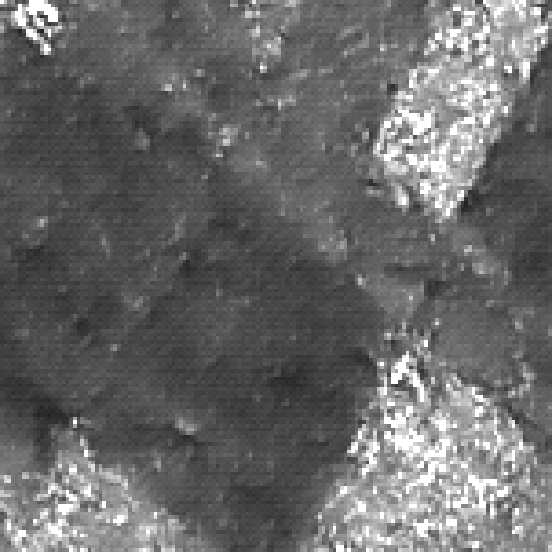}&
     \includegraphics[width=0.15\columnwidth]{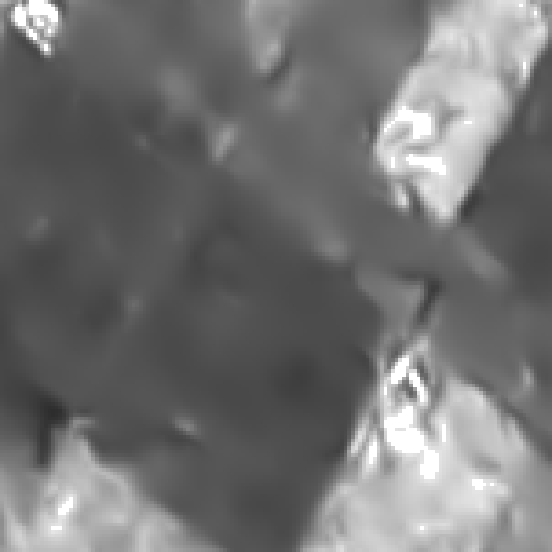}\\
    
     \includegraphics[width=0.15\columnwidth]{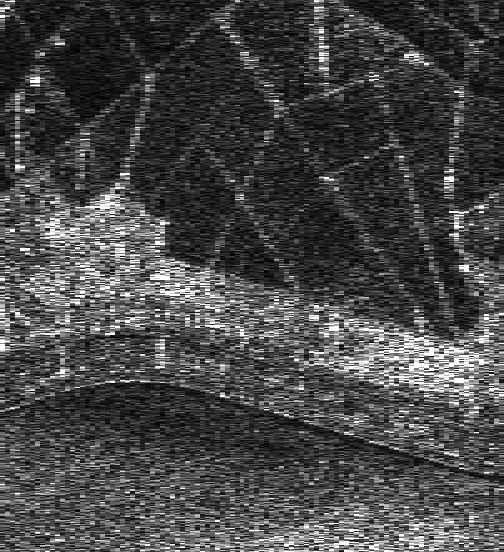}&
     \includegraphics[width=0.15\columnwidth]{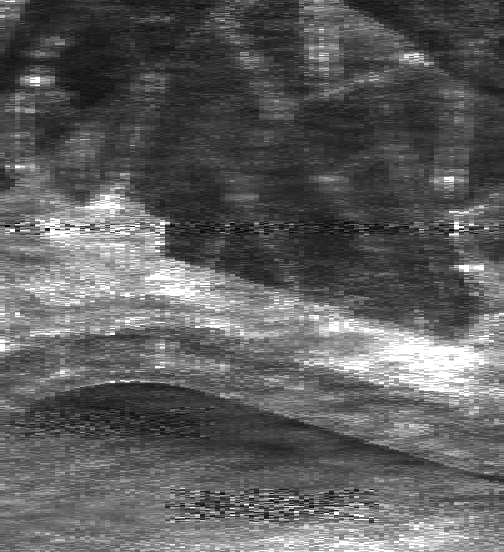}&
     \includegraphics[width=0.15\columnwidth]{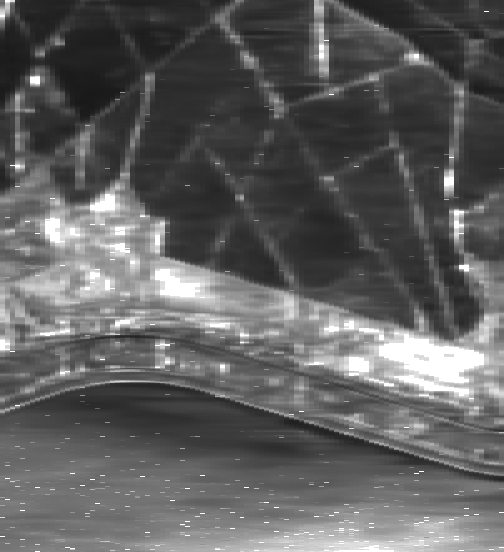}&
     \includegraphics[width=0.15\columnwidth]{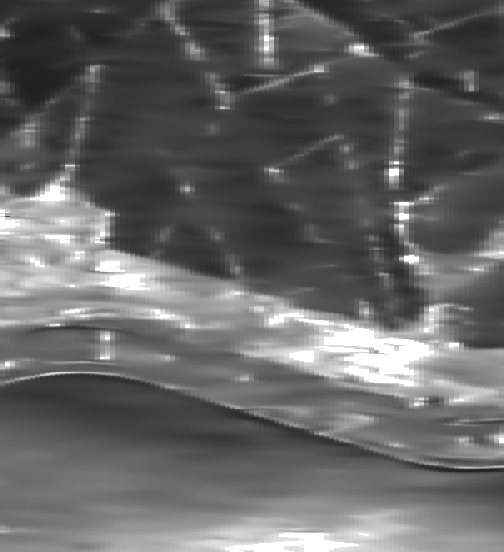}&
     \includegraphics[width=0.15\columnwidth]{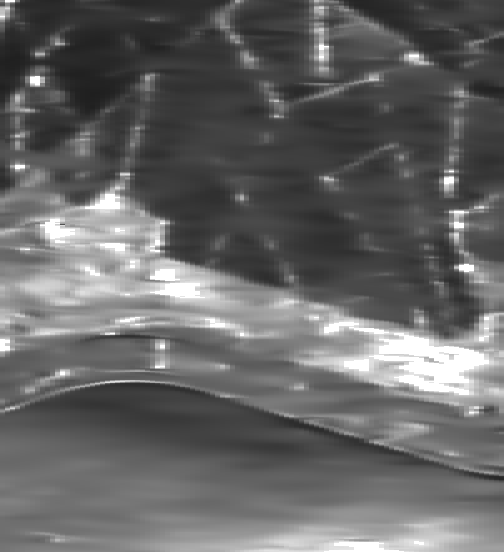}\\

     \includegraphics[width=0.15\columnwidth]{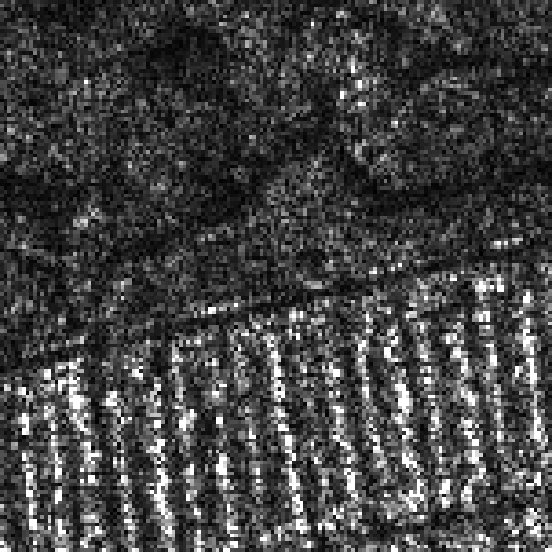}&
     \includegraphics[width=0.15\columnwidth]{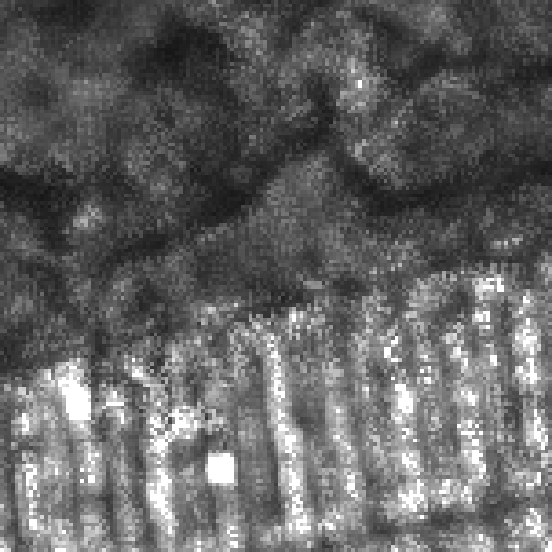}&
     \includegraphics[width=0.15\columnwidth]{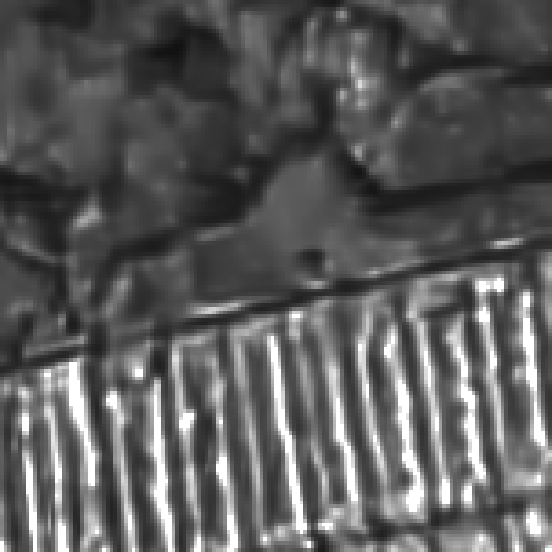}&
     \includegraphics[width=0.15\columnwidth]{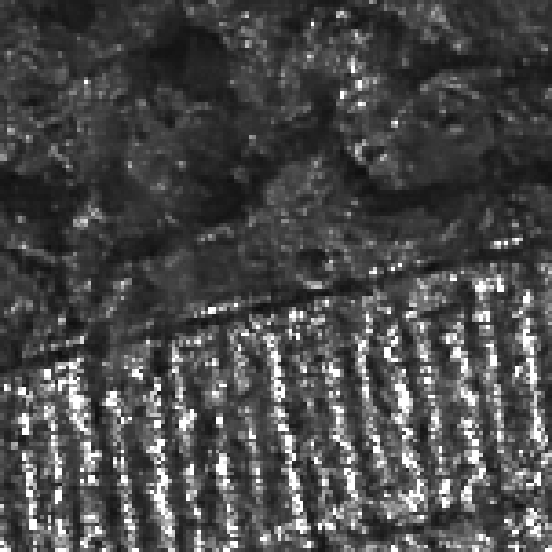}&
     \includegraphics[width=0.15\columnwidth]{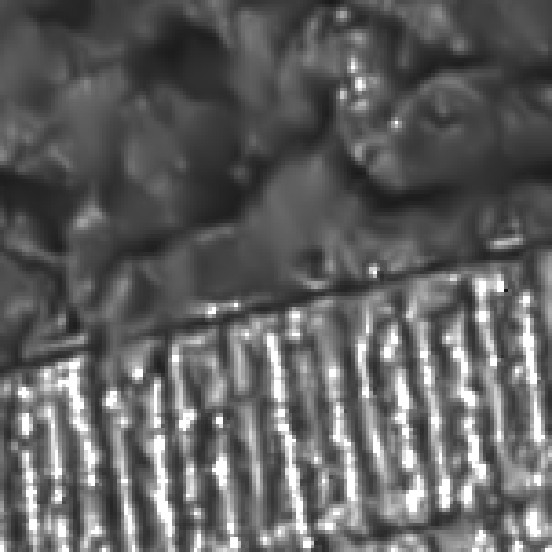}\\

     \end{tabular}
    
    \caption{Close-up of the single-polarization results. From left to right the noisy image and filtering results. From top to bottom RADARSAT-2, BIOMASS and ALOS testing cases.}
    \label{fig:single_comparison_zoom}
\end{figure}

In the single-polarization assessment \prop{} is compared with  MEET-SAR, MuLog, and CL-SAR. The qualitative assessment is reported in Figure \ref{fig:single_comparison}. In this cases only the ability in preserving spatial texture and details can be evaluated. Basically, the comparison between \prop{}, MuLog and MEET-SAR confirms previous observations: MEET-SAR retains spatial details but provides only limited speckle suppression. MuLog effectively balances the trade-off among edge retention, detail preservation, and speckle suppression, achieving the best performance in urban areas. \prop{} offers good spatial detail preservation and edge retention, although slightly inferior to MuLog, especially in urban areas. Finally, CL-SAR struggles to generalize across different datasets, producing images with limited speckle suppression. These characteristics are evident in the close-ups shown in Figure \ref{fig:single_comparison_zoom}.

The quantitative analysis is reported in Table \ref{tab:single_pol}. \prop{} achieves strong performance on almost all metrics: in M-index, MoR, and VoR, \prop{} and MuLog always share the top two positions. Since the M-index combines the ENL with a measure of the residual structures in the ratio image, its good score indicates an effective balance between speckle suppression and texture preservation. The good MoR and VoR values show that the method preserves the statistical properties of the speckle during filtering. \prop{} ranks third in MoI, indicating some difficulty in preserving the mean radiometry of the image (as testified also from the brighter images). MEET-SAR performs worse than \prop{} and MuLog on all metrics except MoI, which confirms its limited speckle suppression. A high MoI score combined with low scores on the other metrics indicates that the filtered image remains close to the original noisy one, with only slight changes in the speckle realization. CL-SAR perform worst on MoR, VoR and MoI indicating a difficulty in preserving speckle noise statistics and handling different dynamics.

\begin{table}[h]
\caption{Quantitative single-polarization despeckling results. The best value for each metric is highlighted in bold.}
\label{tab:single_pol}
\centering
\renewcommand{\arraystretch}{1.5}
\resizebox{\columnwidth}{!}{%
\begin{tabular}{llllll}
\hline
                                                  & Method   & \textit{M}-index $\downarrow$ & MOR $\rightarrow$ 1 & VOR $\rightarrow$ 1 & MOI $\downarrow$ \\ \hhline{=|=====}
\multicolumn{1}{l|}{\multirow{4}{*}{\rotatebox[origin=c]{90}{RADARSAT-2}}}    & 
MEET-SAR & 15.3040 & 0.7947 & 0.4320 &\textbf{0.0352} \\

\multicolumn{1}{l|}{} & MuLog & \textbf{4.7503} & \underline{0.9773} & \underline{0.6071} & \underline{0.1064} \\

\multicolumn{1}{l|}{} & CL-SAR & 11.5405 & 0.9083 & 0.2790 & 0.1182 \\

\multicolumn{1}{l|}{} & PANDA & \underline{6.0333} & \textbf{0.9968} & \textbf{0.7655} & 0.1474  \\
\hhline{=|=====}
\multicolumn{1}{l|}{\multirow{4}{*}{\rotatebox[origin=c]{90}{BIOMASS}}}     & 
MEET-SAR & 25.4885 & 0.8656 & 0.4733 & \textbf{0.0187}  \\

\multicolumn{1}{l|}{} & MuLog & \underline{5.8145} & \underline{1.0059} & \textbf{1.0166} & 0.1390 \\

\multicolumn{1}{l|}{} & CL-SAR & 7.3575 & 1.1368 & 1.5199 & 0.1629 \\

\multicolumn{1}{l|}{} & PANDA & \textbf{5.1059} & \textbf{1.0039} & \underline{1.1812} & \underline{0.1305}\\

\hhline{=|=====}
\multicolumn{1}{l|}{\multirow{4}{*}{\rotatebox[origin=c]{90}{ALOS}}} &
MEET-SAR & 14.3760 & 0.8421 & 0.4212 & \textbf{0.0551} \\
\multicolumn{1}{l|}{}  & MuLog    & \underline{6.7932} & \underline{0.9793} & \textbf{0.7934} & \underline{0.0853} \\
\multicolumn{1}{l|}{}  & CL-SAR & 8.0298 & 1.0578 & 0.6848 & 0.1253  \\
\multicolumn{1}{l|}{}  & PANDA & \textbf{5.0521} & \textbf{1.0072} & \underline{1.3186} & 0.1332  \\
\hhline{=|=====}
\end{tabular}%dual
}
\end{table}

\section{Discussion}
\label{sec:discussion}

The experimental results provide an assessment of the proposed unified despeckling framework across three polarimetric configurations and three independent SAR sensors. Unlike approaches requiring separate models for different input channel dimension, the proposed method employs a single network that can be applied to different polarimetric configurations. 

An important outcome of the experiments is that the proposed approach maintains competitive despeckling performance despite the differences in operating frequency, acquisition characteristics, and available polarimetric information. This behavior supports the effectiveness of the common covariance representation and polarization-aware masking strategy, which allow different polarimetric configurations to be processed within the same network architecture.

In both qualitative and quantitative assessment of full- and dual-polarization, the \prop{} framework demonstrates good speckle suppression, spatial and polarimetric scattering preservation. Indeed, the trace and maximum-eigenvalue analysis are relevant from a polarimetric perspective. The covariance trace describes the total back-scattered power represented by the observed channels, whereas the maximum eigenvalue reflects the dominant component of the covariance spectrum. Consequently, these descriptors characterize complementary aspects of the residual covariance statistics. Their joint evaluation provides a more informative assessment than either descriptor considered independently.
The good performance on the  agreement with the reference trace distribution shows the ability in preserving the spatial properties on intensity channel of the covariance matrix. The good performance on the eigenvalue distribution estimated from cross-polarized channels show ability in handling the different polarization configuration.

In the single-polarization experiments, the proposed method achieves the lowest M-index for ALOS-PALSAR and BIOMASS, while MuLog obtains the best result for RADARSAT-2. The proposed approach also provides the MoR closest to unity for all three acquisitions, indicating favorable mean radiometric consistency according to the adopted ratio-based evaluation. Nevertheless, other methods achieve better results for individual metrics, particularly VoR, MoI. These differences indicate that speckle suppression and residual statistical preservation are not necessarily optimized simultaneously by a single despeckling approach.

A further aspect concerns the generalization capability of the proposed framework. The training references are constructed from a multitemporal stack of co-registered UAVSAR acquisitions collected over a single geographical area. Despite this restricted training source, the network is evaluated on independent acquisitions from three other sensors, including systems operating at different radar frequencies. The absence of sensor-specific fine-tuning demonstrates that the trained model can be transferred to these test datasets without modifying its learned parameters. Nevertheless, the present experiments do not establish universal generalization to all acquisition conditions, geographical environments, or polarimetric measurement characteristics.

\section{Conclusion}
\label{sec:conclusion}

This paper presented a unified deep-learning framework for SAR despeckling across single-, dual-, and full-polarization configurations. The proposed approach employs a  nine-channel covariance representation and polarization-aware masking to accommodate different polarimetric inputs within a single convolutional neural network. By training the network on reference covariance matrices derived from multitemporal UAVSAR observations, the framework avoids the need for separate models or configuration-specific training procedures.

The experimental evaluation, conducted on ALOS PALSAR, RADARSAT-2, and BIOMASS data under three polarimetric configurations for each sensor, showed that the proposed method achieves competitive results against established model-based and deep-learning despeckling approaches. In the single-polarization case, it provided an effective balance between speckle suppression and preservation of the speckle statistics. In the dual- and full-polarization cases, it showed good agreement with the statistical properties of the reference data, preserving the overall polarimetric information. Although some competing methods performed better on specific descriptors for certain sensors, the proposed approach remained among the best-performing methods across all configurations. These results also indicate that relative performance depends on the sensor and on the statistical descriptor considered, with no single method outperforming all others on every criterion.

The main contribution of the proposed framework lies in its ability to process different polarimetric configurations using a single set of learned parameters, without requiring sensor-specific fine-tuning or configuration-dependent retraining. The experimental results demonstrate the feasibility of this unified strategy across sensors operating at different radar frequencies.

\ifCLASSOPTIONcaptionsoff
  \newpage
\fi

% trigger a \newpage just before the given reference
% number - used to balance the columns on the last page
% adjust value as needed - may need to be readjusted if
% the document is modified later
%\IEEEtriggeratref{8}
% The "triggered" command can be changed if desired:
%\IEEEtriggercmd{\enlargethispage{-5in}}

% references section

% can use a bibliography generated by BibTeX as a .bbl file
% BibTeX documentation can be easily obtained at:
% http://mirror.ctan.org/biblio/bibtex/contrib/doc/
% The IEEEtran BibTeX style support page is at:
% http://www.michaelshell.org/tex/ieeetran/bibtex/
\bibliographystyle{IEEEtran}
% argument is your BibTeX string definitions and bibliography database(s)
\bibliography{references}

@inproceedings{he2016,
  author    = {He, Kaiming and Zhang, Xiangyu and Ren, Shaoqing and Sun, Jian},
  title     = {Deep Residual Learning for Image Recognition},
  booktitle = {Proceedings of the IEEE Conference on Computer Vision and Pattern Recognition (CVPR)},
  year      = {2016},
  pages     = {770--778}
}

@article{Vitale2022,
   author = {S. Vitale and G. Ferraioli and V. Pascazio},
   doi = {10.1109/LGRS.2021.3091287},
   journal = {IEEE Geoscience and Remote Sensing Letters},
   title = {Analysis on the Building of Training Dataset for Deep Learning SAR Despeckling},
   volume = {19},
   year = {2022}
}

@ARTICLE{Loic2025,
  author={Denis, Loïc and Dalsasso, Emanuele and Tupin, Florence},
  journal={IEEE Transactions on Geoscience and Remote Sensing}, 
  title={Just Project! Multichannel Despeckling, the Easy Way}, 
  year={2025},
  volume={63},
  number={},
  pages={1-11},
  doi={10.1109/TGRS.2025.3531957}}

@article{Dalsasso2022,
   author = {Emanuele Dalsasso and Loïc Denis and Florence Tupin and Senior Member},
   doi = {10.1109/TGRS.2021.3128621},
   journal = {IEEE TRANSACTIONS ON GEOSCIENCE AND REMOTE SENSING},
   pages = {4704713},
   title = {As If by Magic: Self-Supervised Training of Deep Despeckling Networks With MERLIN},
   volume = {60},
   url = {https://www.ieee.org/publications/rights/index.html},
   year = {2022}
}

@article{Parrilli2012,
   author = {Sara Parrilli and Mariana Poderico and Cesario Vincenzo Angelino and Luisa Verdoliva},
   doi = {10.1109/TGRS.2011.2161586},
   issn = {01962892},
   issue = {2},
   journal = {IEEE Transactions on Geoscience and Remote Sensing},
   pages = {606-616},
   publisher = {IEEE},
   title = {A nonlocal SAR image denoising algorithm based on LLMMSE wavelet shrinkage},
   volume = {50},
   year = {2012}
}

@article{Ferraioli2019,
   author = {Giampaolo Ferraioli and Vito Pascazio and Gilda Schirinzi},
   doi = {10.1109/TGRS.2019.2916465},
   issn = {15580644},
   issue = {10},
   journal = {IEEE Transactions on Geoscience and Remote Sensing},
   pages = {7785-7798},
   publisher = {IEEE},
   title = {Ratio-based nonlocal anisotropic despeckling approach for sar images},
   volume = {57},
   year = {2019}
}

@ARTICLE{Ma2018,
  author={Ma, Xiaoshuang and Wu, Penghai and Wu, Yanlan and Shen, Huanfeng},
  journal={IEEE Journal of Selected Topics in Applied Earth Observations and Remote Sensing}, 
  title={A Review on Recent Developments in Fully Polarimetric SAR Image Despeckling}, 
  year={2018},
  volume={11},
  number={3},
  pages={743-758},
  doi={10.1109/JSTARS.2017.2768059}}

@ARTICLE{Fracastoro2021,
  author={Fracastoro, Giulia and Magli, Enrico and Poggi, Giovanni and Scarpa, Giuseppe and Valsesia, Diego and Verdoliva, Luisa},
  journal={IEEE Geoscience and Remote Sensing Magazine}, 
  title={Deep Learning Methods For Synthetic Aperture Radar Image Despeckling: An Overview Of Trends And Perspectives}, 
  year={2021},
  volume={9},
  number={2},
  pages={29-51},
  doi={10.1109/MGRS.2021.3070956}}

@article{Moreira2013,
   author = {Alberto Moreira and Pau Prats-Iraola and Marwan Younis and Gerhard Krieger and Irena Hajnsek and Konstantinos P. Papathanassiou},
   doi = {10.1109/MGRS.2013.2248301},
   issn = {21686831},
   issue = {1},
   journal = {IEEE Geoscience and Remote Sensing Magazine},
   pages = {6-43},
   publisher = {IEEE},
   title = {A tutorial on synthetic aperture radar},
   volume = {1},
   year = {2013}
}

@article{Deledalle2014,
   author = {Charles Alban Deledalle and Loic Denis and Giovanni Poggi and Florence Tupin and Luisa Verdoliva},
   doi = {10.1109/MSP.2014.2311305},
   issn = {10535888},
   issue = {4},
   journal = {IEEE Signal Processing Magazine},
   pages = {69-78},
   title = {Exploiting patch similarity for SAR image processing: The nonlocal paradigm},
   volume = {31},
   year = {2014}
}

@article{Argenti2013,
   author = {Fabrizio Argenti and Alessandro Lapini and Luciano Alparone and Tiziano Bianchi},
   doi = {10.1109/MGRS.2013.2277512},
   issn = {21686831},
   issue = {3},
   journal = {IEEE Geoscience and Remote Sensing Magazine},
   month = {9},
   pages = {6-35},
   publisher = {Institute of Electrical and Electronics Engineers Inc.},
   title = {A tutorial on speckle reduction in synthetic aperture radar images},
   volume = {1},
   year = {2013}
}

@article{Lee1980,
  author  = {Lee, Jong-Sen},
  title   = {Digital Image Enhancement and Noise Filtering by Use of Local Statistics},
  journal = {IEEE Transactions on Pattern Analysis and Machine Intelligence},
  year    = {1980},
  volume  = {PAMI-2},
  number  = {2},
  pages   = {165--168},
  doi     = {10.1109/TPAMI.1980.4766994}
}

@article{Cloude1997,
  author  = {Cloude, Shane R. and Pottier, Eric},
  title   = {An Entropy Based Classification Scheme for Land Applications of Polarimetric {SAR}},
  journal = {IEEE Transactions on Geoscience and Remote Sensing},
  year    = {1997},
  volume  = {35},
  number  = {1},
  pages   = {68--78},
  doi     = {10.1109/36.551935}
}

@article{Freeman1998,
  author  = {Freeman, Anthony and Durden, Stephen L.},
  title   = {A Three-Component Scattering Model for Polarimetric {SAR} Data},
  journal = {IEEE Transactions on Geoscience and Remote Sensing},
  year    = {1998},
  volume  = {36},
  number  = {3},
  pages   = {963--973},
  doi     = {10.1109/36.673687}
}

@ARTICLE{11028079,
  author={Vitale, Sergio and Ferraioli, Giampaolo and Pascazio, Vito and Deniz, Luis Gomez},
  journal={IEEE Geoscience and Remote Sensing Letters}, 
  title={Enhanced Deep Learning SAR Despeckling Networks Based on SAR Assessing Metrics}, 
  year={2025},
  volume={22},
  number={},
  pages={1-5},
  doi={10.1109/LGRS.2025.3577907}}

@article{Deledalle2015NLSAR,
  author  = {Deledalle, Charles-Alban and Denis, Lo{\"i}c and Tupin, Florence and Reigber, Andreas and J{\"a}ger, Marc},
  title   = {{NL-SAR}: A Unified Nonlocal Framework for Resolution-Preserving {(Pol)(In)SAR} Denoising},
  journal = {IEEE Transactions on Geoscience and Remote Sensing},
  year    = {2015},
  volume  = {53},
  number  = {4},
  pages   = {2021--2038},
  doi     = {10.1109/TGRS.2014.2352555}
}

@article{Deledalle2017MuLoG,
  author  = {Deledalle, Charles-Alban and Denis, Lo{\"i}c and Tabti, Sonia and Tupin, Florence},
  title   = {{MuLoG}, or How to Apply Gaussian Denoisers to Multi-Channel {SAR} Speckle Reduction?},
  journal = {IEEE Transactions on Image Processing},
  year    = {2017},
  volume  = {26},
  number  = {9},
  pages   = {4389--4403},
  doi     = {10.1109/TIP.2017.2713946}
}

@article{Aghababaei2022MEETSAR,
  author  = {Aghababaei, Hossein and Ferraioli, Giampaolo and Vitale, Sergio and Zamani, Roghayeh and Schirinzi, Gilda and Pascazio, Vito},
  title   = {Nonlocal Model-Free Denoising Algorithm for Single- and Multichannel {SAR} Data},
  journal = {IEEE Transactions on Geoscience and Remote Sensing},
  year    = {2022},
  volume  = {60},
  pages   = {1--15},
  doi     = {10.1109/TGRS.2021.3127109}
}

@article{Vitale2021MONet,
  author  = {Vitale, Sergio and Ferraioli, Giampaolo and Pascazio, Vito},
  title   = {Multi-Objective {CNN}-Based Algorithm for {SAR} Despeckling},
  journal = {IEEE Transactions on Geoscience and Remote Sensing},
  year    = {2021},
  volume  = {59},
  number  = {11},
  pages   = {9336--9349},
  doi     = {10.1109/TGRS.2020.3034852}
}

@article{Vitale2023GMonet,
  author  = {Vitale, Sergio and Ferraioli, Giampaolo and Frery, Alejandro C. and Pascazio, Vito and Yue, Dong-Xiao and Xu, Feng},
  title   = {{SAR} Despeckling Using Multiobjective Neural Network Trained With Generic Statistical Samples},
  journal = {IEEE Transactions on Geoscience and Remote Sensing},
  year    = {2023},
  volume  = {61},
  pages   = {1--12},
  doi     = {10.1109/TGRS.2023.3314857}
}

@article{Lu2024Hybrid,
  author  = {Lu, Xialei and Vitale, Sergio and Aghababei, Hossein and Ferraioli, Giampaolo and Pascazio, Vito},
  title   = {Training Supervised Neural Networks for {PolSAR} Despeckling With an Hybrid Approach},
  journal = {IEEE Geoscience and Remote Sensing Letters},
  year    = {2024},
  volume  = {21},
  pages   = {1--5},
  doi     = {10.1109/LGRS.2023.3333671}
}

@article{MestreQuereda2025,
  author  = {Mestre-Quereda, Alejandro and Lopez-Sanchez, Juan M.},
  title   = {Deep Learning-Based Speckle Filtering for Polarimetric {SAR} Images: Application to {Sentinel-1}},
  journal = {IEEE Transactions on Geoscience and Remote Sensing},
  year    = {2025},
  volume  = {63},
  pages   = {1--18},
  doi     = {10.1109/TGRS.2025.3536090}
}

@article{Mendes2025PolSAR2PolSAR,
  author  = {Mendes, Cristiano Ulondu and Dalsasso, Emanuele and Zhang, Yi and Denis, Lo{\"i}c and Tupin, Florence},
  title   = {{PolSAR2PolSAR}: A Semi-Supervised Despeckling Algorithm for Polarimetric {SAR} Images},
  journal = {ISPRS Journal of Photogrammetry and Remote Sensing},
  year    = {2025},
  volume  = {220},
  pages   = {783--798},
  doi     = {10.1016/j.isprsjprs.2025.01.008}
}

@article{Fang2024CLSAR,
  author  = {Fang, Yangtian and Liu, Rui and Peng, Yini and Guan, Jianjun
             and Li, Duidui and Tian, Xin},
  title   = {Contrastive Learning for Real {SAR} Image Despeckling},
  journal = {ISPRS Journal of Photogrammetry and Remote Sensing},
  year    = {2024},
  volume  = {218},
  pages   = {376--391},
  doi     = {10.1016/j.isprsjprs.2024.11.003}
}

@article{Zhang2022DRUNet,
  author  = {Zhang, Kai and Li, Yawei and Zuo, Wangmeng and Zhang, Lei
             and Van Gool, Luc and Timofte, Radu},
  title   = {Plug-and-Play Image Restoration With Deep Denoiser Prior},
  journal = {IEEE Transactions on Pattern Analysis and Machine Intelligence},
  year    = {2022},
  volume  = {44},
  number  = {10},
  pages   = {6360--6376},
  doi     = {10.1109/TPAMI.2021.3088914}
}

@article{Gomez2017MIndex,
  author  = {Gomez, Luis and Ospina, Raydonal and Frery, Alejandro C.},
  title   = {Unassisted Quantitative Evaluation of Despeckling Filters},
  journal = {Remote Sensing},
  year    = {2017},
  volume  = {9},
  number  = {4},
  pages   = {389},
  doi     = {10.3390/rs9040389}
}

@article{Anfinsen2009ENL,
  author  = {Anfinsen, Stian Normann and Doulgeris, Anthony P.
             and Eltoft, Torbj{\o}rn},
  title   = {Estimation of the Equivalent Number of Looks in Polarimetric
             Synthetic Aperture Radar Imagery},
  journal = {IEEE Transactions on Geoscience and Remote Sensing},
  year    = {2009},
  volume  = {47},
  number  = {11},
  pages   = {3795--3809},
  doi     = {10.1109/TGRS.2009.2019269}
}

@ARTICLE{9006944,
  author={Aghababaei, Hossein and Ferraioli, Giampaolo},
  journal={IEEE Geoscience and Remote Sensing Letters}, 
  title={Statistical Indices for Despeckling Evaluation in Multichannel SAR Images}, 
  year={2021},
  volume={18},
  number={2},
  pages={316-320},
  doi={10.1109/LGRS.2020.2973462}}
%
% <OR> manually copy in the resultant .bbl file
% set second argument of \begin to the number of references
% (used to reserve space for the reference number labels box)

% biography section
% 
% If you have an EPS/PDF photo (graphicx package needed) extra braces are
% needed around the contents of the optional argument to biography to prevent
% the LaTeX parser from getting confused when it sees the complicated
% \includegraphics command within an optional argument. (You could create
% your own custom macro containing the \includegraphics command to make things
% simpler here.)
%\begin{IEEEbiography}[{\includegraphics[width=1in,height=1.25in,clip,keepaspectratio]{mshell}}]{Michael Shell}
% or if you just want to reserve a space for a photo:

% \begin{IEEEbiography}{Michael Shell}
% Biography text here.
% \end{IEEEbiography}

% if you will not have a photo at all:
% \begin{IEEEbiographynophoto}{John Doe}
% Biography text here.
% \end{IEEEbiographynophoto}

% insert where needed to balance the two columns on the last page with
% biographies
%\newpage

% \begin{IEEEbiographynophoto}{Jane Doe}
% Biography text here.
% \end{IEEEbiographynophoto}

% You can push biographies down or up by placing
% a \vfill before or after them. The appropriate
% use of \vfill depends on what kind of text is
% on the last page and whether or not the columns
% are being equalized.

%\vfill

% Can be used to pull up biographies so that the bottom of the last one
% is flush with the other column.
%\enlargethispage{-5in}

% that's all folks
\end{document}